\documentclass{pinchcrn}   
                                            
\usepackage{xcolor}
\usepackage{overpic}
\usepackage{amsmath}
\usepackage{bm}
\usepackage{amssymb}
\usepackage{enumitem}

\newcommand{\Dsl}{D\!\!\!\!\slash}

\newcommand{\nsl}{n\!\!\!\slash}

\newcommand{\nbsl}{\bar{n}\!\!\!\slash}

\newcommand{\cc}{c}
\newcommand{\cb}{\bar{c}}
\newcommand{\be}{\begin{equation}}
\newcommand{\ee}{\end{equation}}

\newcommand{\cusp}{\mbox{{\tiny cusp}}}
\newcommand{\bea}{\begin{eqnarray}}
\newcommand{\eea}{\end{eqnarray}}
\newcommand{\bd}{\begin{displaymath}}
\newcommand{\ed}{\end{displaymath}}

\newcommand{\ep}{\varepsilon}

\newcommand{\nn}{\nonumber}

\newcommand{\gsim}{\;\rlap{\lower 3.5 pt \hbox{$\mathchar \sim$}} \raise 1pt \hbox {$>$}\;}
\newcommand{\lsim}{\;\rlap{\lower 3.5 pt \hbox{$\mathchar \sim$}} \raise 1pt \hbox {$<$}\;}

\newcommand{\bc}{\begin{center}}
\newcommand{\ec}{\end{center}}
\newcommand{\ms}{{\overline{\mbox{{\rm MS}}}}}

\def\lapprox{\lower .7ex\hbox{$\;\stackrel{\textstyle <}{\sim}\;$}}
\def\gapprox{\lower .7ex\hbox{$\;\stackrel{\textstyle >}{\sim}\;$}}

\definecolor{NavyBlue}{HTML}{0F75FF}
\definecolor{ForestGreen}{HTML}{00E100}
\definecolor{BrickRed}{HTML}{B80000}

\title{Les Houches Lectures on \\Soft-Collinear Effective Theory}

\author{Thomas Becher}
\affiliation{Albert Einstein Center for Fundamental Physics, Institut f\"ur Theoretische Physik,\\
Universit\"at Bern, Sidlerstrasse 5, CH-3012 Bern, Switzerland}

\begin{document}

\maketitle

\preface

These lectures were part of the Les Houches Summer School {\em Effective Field Theory in Particle Physics and Cosmology}, which took place in July 2017. The school also offered introductory lectures on effective field theory, as well as associated techniques, such as the renormalization group. In this specialized course on Soft-Collinear Effective Theory \cite{Bauer:2000yr,Bauer:2001yt,Beneke:2002ph} I will therefore assume some basic familiarity with effective field theory methods. A reader who is completely new to this area should consult the lectures of Aneesh Manohar and Matthias Neubert at the school, or one of the growing number of lecture notes and books on effective theory \cite{Donoghue:1992dd,Georgi:1994qn,Pich:1998xt,Rothstein:2003mp,Ecker:2005ny,Kaplan:2005es,Grozin:2009an,Petrov:2016azi}.  

Together with Alessandro Broggio and Andrea Ferroglia, I have authored an introductory book on Soft-Collinear Effective Theory \cite{Becher:2014oda}. The construction of the effective theory presented here proceeds along similar lines and follows the same philosophy as the book (which in turn is guided by \cite{Beneke:2002ph}). However, to keep things interesting for myself and hopefully useful for the reader, the physics examples are completely disjoint from the ones in the book. In the book, we focused on threshold resummation and transverse momentum resummation, while I am discussing jet processes and soft-photon physics in the present lecture notes. In the book, every single computation is spelled out step by step. I needed to move at a somewhat quicker pace to cover the relevant material in these lectures, but will point to the detailed derivations in the book in those cases where I won't be able present them here.

\tableofcontents

\maintext

\chapter{Introduction}

Soft-Collinear Effective Theory (SCET) \cite{Bauer:2000yr,Bauer:2001yt,Beneke:2002ph} is the effective field theory for processes with energetic particles such as jet production at high-energy colliders. A typical two-jet process is depicted in Fig.\ \ref{twojet}. It involves sprays of energetic particles along two directions with momenta $p_J$ and $p_{\bar{J}}$, accompanied by soft radiation with momentum $p_s$. Typically such processes involve a scale hierarchy
\begin{equation}\label{scales}
Q^2 = (p_J + p_{\bar{J}} )^2 \gg p_J^2 \sim  p_{\bar{J}}^2  \gg p_s^2\,.
\end{equation}
In SCET, the physics associated with the hard scale $Q^2$ is integrated out and absorbed into Wilson coefficients of effective-theory operators, much in the same way that a heavy particle is integrated out when constructing a low-energy theory for light particles only. SCET involves two different types of fields, collinear and soft fields to describe the physics associated with the two low-energy scales $p_J^2$ and $p_s^2$.

The result of a SCET analysis of a jet cross section is often a factorization theorem of the schematic form
\begin{equation}\label{factSchem}
\sigma  = H \cdot J \otimes \bar{J} \otimes S\,.
\end{equation}
The hard function $H$ encodes the physics at the scale $Q^2$, while the jet functions $J$ and $\bar{J}$ depend on the jet scales $p_J^2$  and $p_{\bar{J}}^2$, respectively,  and the soft function $S$ describes the physics at the soft momentum scale. Depending on the observable under consideration, the jet and soft functions are convoluted or multiplied together, as indicated by the $\otimes$ symbol. We will derive several such factorization theorems during the course of these lectures.

The theorem \eqref{factSchem} is obtained after expanding in the ratios of the scales \eqref{scales} and holds at leading power in the expansion. Its main virtue is scale separation -- the property that each of the functions in \eqref{factSchem} is only sensitive to a single scale. The individual functions furthermore fulfill renormalization group (RG) equations. As was demonstrated in Matthias Neubert's lecture \cite{NeubertLecture}, by solving RG equations one can resum the large perturbative logarithms. To do so one evaluates each component function in \eqref{factSchem} at its natural scale and then evolves them to a common reference scale where they are combined. This resums logarithms such as $\alpha^{n} \ln^m(Q^2/p_J^2)$ with $m\leq 2n$, which can spoil the standard perturbative expansion of the cross section. A characteristic feature of jet processes is the presence of double logarithms, $m=2n$, which are also called Sudakov logarithms. They are the result of an interplay of soft and collinear physics.

In certain cases, the soft or collinear scales can be so low that a perturbative expansion becomes unreliable, even after resumming the large logarithms. Factorization theorems such as \eqref{factSchem} remain useful also in this situation, because they allow one to separate perturbative from non-perturbative physics. In fact, every high-energy hadron-collider computation involves non-perturbative Parton-Distribution Functions (PDFs). Having a factorization theorem which separates them from the hard scattering process is crucial to be able to make predictions.

{\footnotesize
\begin{figure}[t!]        
  \begin{center}             
\begin{overpic}[scale=0.7]{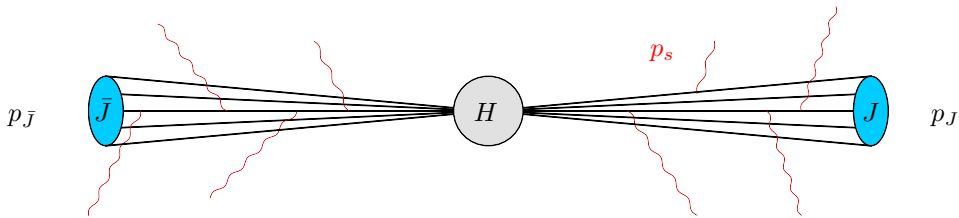}
\put(105,12){$p_J$}
\put(-10,12){$p_{\bar{J}}$}
\put(48,12){$H$}
\put(0.7,12){$\bar{J}$}
\put(96.4,12){$J$}
\put(70,20){$\textcolor{red}{p_s}$}
\end{overpic}
\end{center}
\vspace{-0.3cm}
\caption{Sketch of a two jet process.}   
\label{twojet}   
\end{figure}

}

Traditionally, factorization theorems were derived purely diagrammatically, see \cite{Collins:1989gx,Sterman:1995fz,Sterman:1996uf,Collins:2011zzd} for an introduction to these methods. SCET is of course based on the same physics as the diagrammatic techniques and there is a close relation between the two approaches. An advantage of SCET is that effective theory provides an operator formulation of the low-energy physics, which simplifies and systematizes the analysis. This is especially important for complicated problems such as the factorization of power suppressed contributions. Via the RG equations, SCET also provides a natural framework to perform resummations.

Compared to traditional effective field theories such as Fermi theory, SCET involves several complications. First of all, we cannot simply integrate out particles: quarks and gluons are still present in the low-energy theory. Instead, one splits the fields into modes
\begin{align}
	&\,   H \;\;\phantom{+}\; J \;\;\phantom{+}\; \bar{J}\;\; \phantom{+}\;S \nonumber \\
\phi  \,=\, &\, \phi_h + \phi_c + \phi_{\bar{c}} + \phi_s\,
\end{align}
containing the contributions of the different momentum regions of the quark or gluon fields $\phi \in \{ q, A_\mu^a\}$. The hard mode $\phi_h$, which describes contributions where the particles are off shell by a large amount, is then ``integrated out'', while the low-energy modes become the fields of the effective theory. The term ``integrating out'' refers to a path-integral formalism introduced by Wilson, where one would integrate over the field $\phi_h$ since it does not appear as an external state. This is, however, not how things are done in practice. Instead, one starts by writing down the most general effective Lagrangian with the low-energy fields. One then adjusts the couplings of the different terms in the Lagrangian to reproduce the contribution of the hard momentum region. These ``couplings'' are also called {\em Wilson coefficients} and the process of adjusting them to reproduce the full theory result is called {\em matching}, see e.g.\ \cite{ManoharLecture}. An important and nontrivial element of the analysis is to identify the relevant momentum modes for the problem at hand, which are the degrees of freedom of the effective theory. This is done by analyzing full theory diagrams and provides the starting point of the effective theory construction.

Not only does SCET contain several different fields for each QCD particle, a second complication is that the different momentum components of the fields scale differently. Momentum components transverse to the jet direction are always small, but the components along the jet directions are large. To perform a derivative expansion of the effective Lagrangian, one therefore needs to split the momenta into different components. This is done by introducing reference vectors $n^\mu \propto p_J^\mu$ and $\bar{n}^\mu \propto p_{\bar{J}}^\mu$  in the directions  of the two jets. The fact that the momentum components of the collinear particles along the jet are unsuppressed leads to a final complication, namely that one can write down operators with an arbitrary number of such derivatives. One way to take all these operators into account is to make operators nonlocal along the corresponding light-cone directions, as we will explain later. 

\chapter{Warm up: soft effective theory}
\label{set}

For the reasons discussed in the introduction, the construction of SCET is technically involved. We will address all of the complications, but due to their presence it will take some time to set up the effective Lagrangian. Rather than immediately diving into technicalities, I would like to first discuss a simpler example where only some of the difficulties are present and where we can obtain physics results a bit quicker.  Therefore, before turning to jet processes, I want to start with electron-electron scattering in QED,  as depicted in Fig.\ \ref{eeScatt}. 

To keep things simple, I will assume that the electron energies are of the order of the electron mass $m_e$.  The electrons can thus be relativistic but not ultra-relativistic (having them ultra-relativistic would bring us back into the realm of SCET). Instead of SCET, we then deal with SET, i.e.\ soft effective field theory. We will use SET to obtain a classic factorization theorem for soft photon radiation in QED. This seminal result was originally derived by Yennie, Frautschi and Suura in 1961 using diagrammatic methods \cite{Yennie:1961ad}. All the steps in our derivation are also valid in QCD and later in the course we will use our results to analyze soft-gluon effects in jet processes.

\section{Soft photons in electron-electron scattering}

Note that we cannot avoid the presence of soft photons in QED processes. No matter how good our detectors are, photons with energies below some threshold will always go unnoticed and since it costs little energy to produce them, any QED final state will always include soft photons. Indeed, trying to compute higher-order corrections to scattering processes without accounting for their presence leads to divergent results for cross sections. The divergence signals that completely exclusive QED cross sections are not physical, which was understood very early on \cite{Bloch:1937pw,Kinoshita:1962ur,Lee:1964is}. When we talk about electron-electron scattering, we really measure the inclusive process
\begin{equation}\label{eescattKin}
e^-(p_1) + e^-(p_2)  \to e^-(p_3) +e^-(p_4) + X_s(q_s)\,,
\end{equation}
where $X_s$ is any state with an arbitrary number of soft photons which carry the total momentum $q_s$. The constraints imposed on the soft momenta depend of course on the detailed experimental setup, but for our purposes it will be sufficient to assume that the total energy fulfills $E_s \ll m_e$. We will now analyze the process \eqref{eescattKin} up to terms suppressed by powers of the expansion parameter $\lambda = E_\gamma / m_e$. Given the large scale hierarchy, effective-field theory methods should be useful to analyze this situation. 

So what is the effective Lagrangian for soft photons with $E_\gamma \ll m_e$ by themselves? As always, it is useful to organize the operators in the effective Lagrangian by their dimension
\begin{equation}\label{Lgamma}
\mathcal{L}_{\rm eff}^\gamma = \mathcal{L}_{4}^\gamma + \frac{1}{m_e^2}  \mathcal{L}_{6}^\gamma  + \frac{1}{m_e^4} \mathcal{L}_{8}^\gamma\,
\end{equation}
because the coefficients of the higher-dimensional operators involve inverse powers of the large scale, which is $m_e$ in our case. The contributions of these operators will therefore be suppressed by powers of $\lambda$. The leading Lagrangian only involves a single term
\begin{equation}
\mathcal{L}_{4}^\gamma =  - \frac{1}{4} F_{\mu\nu} F^{\mu\nu}\,,
\end{equation}
whose coefficient can be adjusted to the canonical value by rescaling the photon field. The leading-power effective-field-theory Lagrangian is therefore simply the one for free photons. This makes sense, since the effective theory is obtained by integrating out the massive particles which leaves only the photons. Integrating out the electrons does induce higher-power operators which describe photon-photon interactions. While we will not need them, it is an interesting exercise to analyze these higher-power terms; the first nontrivial ones arise at dimension 8, see e.g.\ \cite{Grozin:2009an,Burgess:2007pt}.

{\footnotesize
\begin{figure}[t!]       
\begin{center}              
\begin{overpic}[scale=0.8]{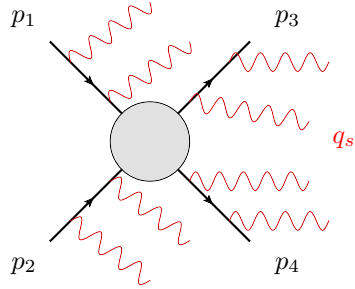}
\put(-12,92){$p_1$}
\put(-12,6){$p_2$}
\put(80,92){$p_3$}
\put(80,6){$p_4$}
\put(100,50){$\textcolor{red}{q_s}$}
\end{overpic}
\end{center}
\caption{Electron-electron scattering, including soft-photon radiation.}   
\label{eeScatt}  
\end{figure}
}

However, $\mathcal{L}_{\rm eff}^\gamma$ is by itself not sufficient. While the energy of the soft radiation is too small to produce additional electron-positron pairs, we do need to include the incoming and outgoing electrons in the effective theory. (Due to fermion number conservation they remain even as $E_\gamma \to 0$.) Consider, as in Fig.\ \ref{outel}, an outgoing electron with momentum $p^\mu = m_e v^\mu$. The momentum on internal fermion lines is $p+q$, where $q$ is a soft photon momentum. We can expand the internal fermion propagators in the small momentum $q$. Neglecting higher order terms suppressed by $q^\mu/m_e$, we find
\begin{equation}
\begin{aligned}\label{propex}
\Delta_F(p+q) &= i \frac{p\!\!\! / + q \!\!\! / + m_e}{(p+q)^2- m_e^2 + i0}  =  i \frac{p\!\!\! / + m_e}{2 p\cdot q+ i0} = \frac{v\!\!\! / +1}{2} \frac{i}{v\cdot q+ i0}\\
& \equiv P_v \,\frac{i}{v\cdot q+ i0}\,,
\end{aligned}
\end{equation}
where we introduced the projection operator
\begin{equation}
P_v = \frac{1+ v\!\!\! /}{2}\,,
\end{equation}
which has the properties
\begin{align}
v\!\!\!/ \, P_v  &= P_v \,, & P_v^2 &= P_v\,,  &  P_v\, \varepsilon\!\!\!/\, P_v &= P_v  \, \varepsilon \cdot v\,.
\end{align}
Using these, the outgoing-leg part of the diagram in Fig.\ \ref{outel} simplifies to
\begin{equation}\label{diagex}
\bar{u}(p)\,P_v\, \frac{i}{v\cdot q} (-i e\, \varepsilon \cdot v) \,P_v\, \frac{i}{v\cdot q'} (-i e\, \varepsilon' \cdot v) \dots\,.
\end{equation}
This form of the expanded soft emissions is well known and called the {\em eikonal approximation}.

{\footnotesize
\begin{figure}[t!]        
\begin{center}              
\begin{overpic}[scale=0.8]{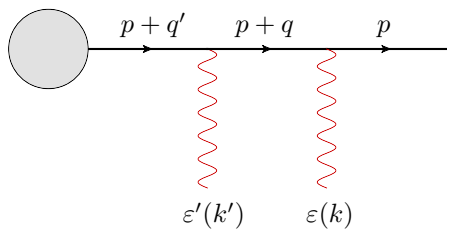}
\put(40,-7){$\varepsilon'(k')$}
\put(68,-7){$\varepsilon(k)$}
\put(26,36){$p+q'$}
\put(52,36){$p+q$}
\put(84,36){$p$}
\end{overpic}\\[10pt]
\end{center}
\caption{Soft emissions from an outgoing electron. Note that $q'=k+k'$ and $q=k$.}   
\label{outel}    
\end{figure}
}

Can we obtain the expanded expression from an effective Lagrangian? This should be possible, we just need to view the expanded propagator \eqref{propex} as the propagator in the effective theory and the emissions in the expanded diagram must be resulting from a Feynman rule $-i e\, v^\mu$ for the electron-photon vertex. So we already know the Feynman rules of the effective theory and just need to write down a Lagrangian which produces them! Consider 
\begin{equation}\label{Lv}
\mathcal{L}_{\rm eff}^v = \bar{h}_v(x)\, i v\cdot D\, h_v(x) \,,
\end{equation}
where $h_v$ is an auxiliary fermion field which fulfills $P_v \,h_v = v\!\!\!/ h_v= h_v$. (Such a field can be obtained by multiplying a regular fermion field with $P_v$.) As usual, the propagator can be obtained by inverting the quadratic part of the Lagrangian in Fourier space and multiplying by $i$. This indeed yields $i/(v\cdot q)$ as in \eqref{propex}. The factor of $P_v$ arises because the external spinor of the auxiliary field $h_v$ includes such a factor due to the property $P_v \,h_v =  h_v$ of the field. Thanks to the projection property $P_v^2=P_v$ a single power of this matrix on the fermion line is sufficient. Also the photon vertex comes out correctly. Inserting $D_\mu = \partial_\mu + i e A_\mu$ into \eqref{Lv}, the vertex Feynman rule is $- i e v^\mu$ so that the Lagrangian \eqref{Lv} correctly reproduces the eikonal expression \eqref{diagex} obtained by expanding the original QED diagram. Note that the propagator of the field $h_v$ only has a single pole in the energy corresponding to the fermion. The anti-fermion pole has been lost in the expansion \eqref{propex}. This is perfectly fine since the field $h_v$ describes a fermion close to its mass-shell with momentum $m_e v^\mu + q^\mu$, where $q^\mu$ is a soft-photon momentum. In this situation anti-fermions cannot arise as external particles and their virtual effects can be absorbed into the Wilson coefficients of the effective theory.

Our construction of \eqref{Lv} highlights the close connection between diagrammatic methods and effective field theory: we constructed the effective Lagrangian in such a way that it reproduces the expansion of the full-theory diagram in \eqref{diagex}. We will follow the same strategy when setting up the SCET Lagrangian below. The astute reader will have recognized \eqref{Lv} as the Lagrangian of Heavy Quark Effective Theory (HQET), covered in Thomas Mannel's lectures \cite{MannelLecture} which contain a path-integral derivation of the same Lagrangian.

The field $h_v$ cannot describe other fermion lines in the process \eqref{eescattKin} which have different velocities. To account for all four fermion lines in the process, we need to include four auxiliary fermion fields so that the full effective Lagrangian takes the form
\begin{equation}
\mathcal{L}_{\rm eff}= \sum_{i=1}^4 \bar{h}_{v_i}(x)\, i v_i\cdot D\, h_{v_i}(x) - \frac{1}{4} F_{\mu\nu} F^{\mu\nu} + \Delta \mathcal{L}_{\rm int}\,,
\end{equation}
where the velocity vectors are given by $v_i^\mu = p_i^\mu/m_e$. This Lagrangian has some features which will also be present in SCET. First of all, $\mathcal{L}_{\rm eff}$ depends on reference vectors along the direction of the energetic particles. In SCET, we will deal with jets of energetic massless particles and the reference vectors will be light-cone vectors along the directions of the jets instead of the velocity vectors. Secondly, we need different fields to represent the electrons along the different directions in the effective theory, while all of these were described by a single field in QED. The same will be true in SCET. In the present case, the different fields are modes of the full-theory fermion field, which live in small momentum regions around the reference momenta $m_e v_i^\mu$. 

{\footnotesize
\begin{figure}[t!]        
\begin{center}              
\begin{tabular}{ccccc}
\includegraphics[height=0.12\textwidth]{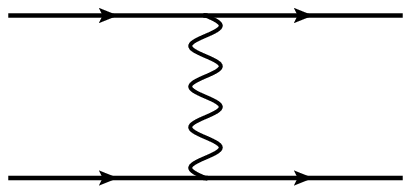}& & \includegraphics[height=0.12\textwidth]{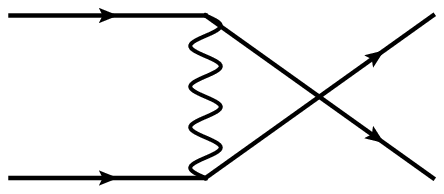}
\end{tabular}
\end{center}
\caption{QED tree-level diagrams for the matching computation needed to extract the Wilson coefficient \eqref{Lint}.}   
\label{eediag}    
\end{figure}

}

What remains is to write down the interaction terms. These have the form
\begin{equation}\label{Lint}
\Delta \mathcal{L}_{\rm int} = C_{\alpha\beta\gamma\delta}(v_1,v_2,v_3,v_4,m_e) \, h^\alpha_{v_1}(x)\, h^\beta_{v_2}(x) \,\bar{h}^\gamma_{v_3}(x) \, \bar{h}^\delta_{v_4}(x)\,.
\end{equation}
More elegantly we could write
\begin{equation}\label{LintAlt}
\Delta \mathcal{L}_{\rm int} = \sum_i C_{i}(v_1,v_2,v_3,v_4,m_e) \, \bar{h}_{v_3}(x) \Gamma_i h_{v_1}(x)\,
 \bar{h}_{v_4}(x) \Gamma_i h_{v_2}(x) \,,
\end{equation}
with a basis of Dirac matrices $ \Gamma_i$ since $\Delta \mathcal{L}_{\rm int}$ has to be a scalar. However, the less elegant form \eqref{Lint}, in which the Wilson coefficients depend on the Dirac indices $\alpha$, $\beta$, $\gamma$, $\delta$ of the fields, will be convenient to perform the matching. In principle we could also write down interaction terms involving only two fields such as 
\begin{equation}\label{LintTwo}
\Delta \mathcal{L}_{\rm int}'  = C_{\alpha\beta}(v_1,v_3) \, h^\alpha_{v_1}(x) \bar{h}^\beta_{v_3}(x)\,,
\end{equation}
but when performing a matching computation one would find that their Wilson coefficients $C_{\alpha\beta}$ are zero if the velocities are different, since the corresponding operator would describe a process in which a fermion spontaneously changes its velocity, which violates momentum conservation. Adding an additional photon field to the operator \eqref{LintTwo} would allow for very small velocity changes, but $\mathcal{O}(1)$ changes are again impossible, if the only particles in the theory are soft photons. This leaves \eqref{Lint} as the simplest nontrivial interaction term. Of course, one could also write interactions terms with covariant derivatives or more fields, but these are higher-dimensional operators, whose contributions are suppressed by powers of the electron mass as in \eqref{Lgamma}.

The leading-power effective Lagrangian is thus complete. All that is left is to determine the Wilson coefficients $C_{\alpha\beta\gamma\delta}(v_1,v_2,v_3,v_4)$ in the interaction term. To do so, we should  compute the same quantity in QED and in the effective theory and then adjust the Wilson coefficient to reproduce the QED result. The simplest quantity we can use to do the matching is the amputated on-shell Green's function for $e^-(p_1) + e^-(p_2)  \to e^-(p_3) +e^-(p_4)$. The relevant QED diagrams are shown in Fig.\ \ref{eediag}. In the effective theory, only the interaction Lagrangian \eqref{Lint} contributes and the result is directly equal to the Wilson coefficient
\begin{equation}
\raisebox{-0.5cm}{\includegraphics[height=0.08\textwidth]{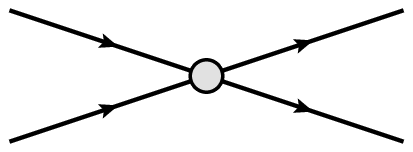}} \;\; = \;\; C_{\alpha\beta\gamma\delta}(v_1,v_2,v_3,v_4,m_e) \,.
\end{equation}
To reproduce the QED result, the Wilson coefficient must be set equal to the on-shell QED Green's function (which is the same as the scattering amplitude, up to the external spinors). At the moment, we are only discussing tree-level matching but the same simple relation also holds at loop level in dimensional regularization. The reason is that all loop corrections to the on-shell amplitude vanish in the effective theory because they are given by scaleless integrals. This is the case since all the photon momenta were set to zero and the electron mass is no longer present in the low-energy theory. This makes it very convenient to use such amplitudes for the matching and shows that the Wilson coefficient has a simple interpretation.

Our effective theory factorizes low- and high-energy physics: the hard scattering of the electrons is part of the Wilson coefficient, which depends on the high-energy scale $m_e$, while the low-energy diagrams in the effective theory only depend on photon-energy scales. We can obtain a very elegant form of the low-energy matrix element by introducing the {\em Wilson line}
\begin{equation}\label{wilson}
S_i(x) = \exp\left[ -i e \int_{-\infty}^0 \!ds\, v_i \cdot A(x+s v_i)\right]\,.
\end{equation}
One way to obtain such a Wilson line is to add a point-like source which travels along the path $y^\mu(s) = x^\mu+s v_i^\mu$ to the Lagrangian of electrodynamics. This is indeed how the outgoing electrons behave: since their energy is much larger than the photon energies, they travel without recoiling when emitting photons. 

To see that \eqref{wilson} reproduces the emission pattern of an incoming electron, let us take the matrix element with a photon in the final state. To obtain it, we can expand the Wilson line in the coupling and since we are taking this matrix element in a free theory (see \eqref{Lgamma}) only the first-order term gives a nonvanishing contribution:
\begin{equation}\label{WilsonMatrix}
\begin{aligned}
\langle \gamma(k) | S_i(0) |0\rangle &= -i e \int_{-\infty}^0 \!ds\, v_i^\mu \langle \gamma(k) | A_\mu(s v_i) |0\rangle \\
&= -i e \int_{-\infty}^0 \!ds\, v_i\cdot \varepsilon(k) e^{i s v_i\cdot k}  
=  e \frac{  v_i\cdot \varepsilon(k) }{-v_i \cdot k + i 0}\,.
\end{aligned}
\end{equation}
We indeed reproduce the eikonal structure  \eqref{diagex} we found expanding the diagram. To ensure the convergence of the integral in the second line of \eqref{WilsonMatrix}  at $s=-\infty$, the exponent $v_i\cdot k$ must have a negative imaginary part, which amounts to the $+ i 0$ prescription in the eikonal propagator. Analogously to \eqref{wilson}, we can also define a Wilson line describing the radiation of an outgoing particle
\begin{equation}\label{wilsonOut}
\bar{S}^\dagger_i(x) = \exp\left[ -i e \int_0^{\infty} \!ds\, v_i \cdot A(x+s v_i)\right]\,.
\end{equation}
We define the dagger of the Wilson line since the outgoing particle is produced by the conjugate field. The matrix element of $\bar{S}_i(x)$ is the same eikonal expression as \eqref{WilsonMatrix}, but with a $- i 0$ prescription.

What is important in the following is that the Wilson line fulfills the equation
\begin{equation}\label{WilsonDiff}
v_i \cdot D \, S_i(x) = 0\,,
\end{equation}
and another way of introducing the object \eqref{wilson} is to define it as the solution to this differential equation. The reader is invited to check that expression \eqref{wilson} indeed fulfills \eqref{WilsonDiff}. An explicit solution to this exercise can be found in Appendix D of \cite{Becher:2014oda}. 

Let us now perform a field redefinition by writing the fermion fields of the incoming fermion fields in the form
\begin{equation}\label{decoupl}
h_{v_i}(x) = S_i(x)\, h_{v_i}^{(0)}(x) \,,
\end{equation}
as a Wilson line along the corresponding direction times a new fermion field $h_{v_i}^{(0)}(x)$. The fermion Lagrangian then takes the form
\begin{equation}
\begin{aligned}
\bar{h}_{v_i}(x)\, i v_i\cdot D\, h_{v_i}(x) &= \bar{h}^{(0)}_{v_i}(x)\, S^\dagger_i(x) \, i v_i\cdot D\,S_i(x)\, h^{(0)}_{v_i}(x) \\
& =  \bar{h}^{(0)}_{v_i}(x)\, S^\dagger_i(x) \,S_i(x) \, i v_i\cdot \partial\,\, h^{(0)}_{v_i}(x) \\
&=  \bar{h}^{(0)}_{v_i}(x) \, i v_i\cdot \partial\,\, h^{(0)}_{v_i}(x)\,.
\end{aligned}
\end{equation}

The field $h_{v_i}^{(0)}(x)$ is a free field; we were able to remove the interactions with the soft photons using the {\em decoupling transformation} \eqref{decoupl}. For the fields describing the outgoing fields, the decoupling is performed using $\bar{S}_i(x)$.\footnote{If we have in- and outgoing fields along the same direction, it is no longer clear how to decouple the soft radiation. Indeed, if we consider nearly forward scattering, the effective theory we formulated is no longer appropriate. One needs to include and resum the effects associated with the Coulomb potential between the two electrons in the process. Similar effects arise in QCD and go under the name of Coulomb or Glauber gluons.} The same method is used in SCET to decouple soft gluons \cite{Bauer:2001yt} as we will see in the next chapter.

While the Wilson lines cancel in the fermion Lagrangian, they are present in the interaction Lagrangian \eqref{LintAlt} which now takes the form
\begin{equation}
\Delta \mathcal{L}_{\rm int} = \sum_i C_{i}(v_1,v_2,v_3,v_4) \; \bar{h}^{(0)}_{v_3}\, \bar{S}^\dagger_3\, \Gamma_i \,S_1\,h^{(0)}_{v_1}\;\;
 \bar{h}^{(0)}_{v_4} \,\bar{S}^\dagger_4\, \Gamma_i \,S_2\, h^{(0)}_{v_2}\,
\end{equation}
so that we end up with Wilson lines along the directions of all particles in the scattering process.

As a final step, we now use our effective theory to compute the scattering amplitude for $\mathcal{M}(e^-(p_1) + e^-(p_2)  \to e^-(p_3) +e^-(p_4) +X_s(k))$, where the final state contains $n$ photons, $X_s(k)= \gamma(k_1) + \gamma(k_2) + \dots \gamma(k_n)$. 
Since the photons no longer interact with the fermions after the decoupling, the relevant matrix element factorizes into a fermionic part times a photonic matrix element. Using the form \eqref{LintAlt} of the interaction Lagrangian, the amplitude is given by
\begin{align}\label{softfact}
\mathcal{M} &= \sum_i C_i\, \bar{u}(v_3) \,\Gamma_i \, u(v_1)\; \bar{u}(v_4) \,\Gamma_i \, u(v_2) \; \langle X_s(k) |  \bar{S}^\dagger_3 \,S_1 \, \bar{S}^\dagger_4\, S_2 | 0 \rangle  \nonumber \\
& =   \mathcal{M}_{ee}  \, \langle X_s(k) |  \bar{S}^\dagger_3 \,S_1 \, \bar{S}^\dagger_4\, S_2 | 0 \rangle\,,
\end{align}
where we have used in the second line that the Wilson coefficient times the spinors is simply the amplitude $\mathcal{M}_{ee} = \mathcal{M}(e^-(p_1) + e^-(p_2)  \to e^-(p_3) +e^-(p_4) )$ for the process without soft photons. So we have shown that the amplitude factorizes into an amplitude without soft photons times a matrix element of Wilson lines. Analogous statements hold for soft gluon emissions in QCD, except that the Wilson lines will be matrices in color space and one has to keep track of the color indices. We can square our factorized amplitude to obtain the cross section, which takes the form
\begin{equation}\label{fact}
\sigma = \mathcal{H}(m_e, \{ \underline{v} \} )\, \mathcal{S}(E_s,  \{ \underline{v} \})\,,
\end{equation}
where the hard function $\mathcal{H}$ is the cross section for the process without soft photons, 
\begin{equation}\label{hardF}
\mathcal{H}(m_e, \{ \underline{v} \} ) = \frac{1}{2E_1 2 E_2 |\vec{v}_1 - \vec{v}_2|} \frac{d^{3}p_3}{(2\pi)^{3} 2E_3}  \frac{d^{3}p_4}{(2\pi)^{3} 2E_4}   |\mathcal{M}_{ee} |^2  (2\pi)^4 \delta^{(4)}( p_1+p_2 -p_3-p_3)\,,
\end{equation}
while the soft function $\mathcal{S}$ is the Wilson line matrix element squared, together with the phase-space constraints on the soft radiation,
\begin{equation}\label{softF}
\mathcal{S}(E_s,  \{ \underline{v} \}) =  \int\limits_{X_s}\hspace{-0.55cm}\sum \left| \langle X_s |  \bar{S}^\dagger_3 \,S_1 \, \bar{S}^\dagger_4\, S_2 | 0 \rangle \right|^2 \theta(E_s - E_{X_s})\,.
\end{equation}
The sum and integral symbol indicates that one has to sum over the different multi-photon final states and integrate over their phase space. Note that both the hard and soft functions depend on the directions $\{ \underline{v} \} = \{ v_1 , \dots, v_4 \}$ of the electrons. For simplicity, we only constrain the total soft energy; the constraints in real experiments will of course be more complicated. To obtain \eqref{fact} we expanded the small soft momentum out of the momentum conservation $\delta$ function
\begin{equation} 
\delta^{(4)}( p_1+p_2 -p_3-p_3-k) = \delta^{(4)}( p_1+p_2 -p_3-p_3) + \mathcal{O}(\lambda)\,,
\end{equation}
which is then part of the hard function $\mathcal{H}$ in \eqref{hardF}.

The Wilson-line matrix elements such as \eqref{softF} which define the soft functions have a very interesting property in QED: they exponentiate, 
\begin{equation}
\mathcal{S}(E_s,  \{ \underline{v} \}) = \exp\left[  \frac{\alpha}{4\pi} \,S^{(1)}(E_s,  \{ \underline{v} \}) \right]\,,
\end{equation}
so that the all-order result is obtained by exponentiating the first-order result. We will not derive this formula here, but the key ingredient in the derivation is the eikonal identity
\begin{equation}\label{eik}
\frac{1}{ v\cdot k_1 \, v\cdot (k_1+k_2 )} + \frac{1}{ v\cdot k_2 \, v\cdot (k_1+k_2 )} = \frac{1}{v\cdot k_1} \frac{1}{v\cdot k_2} \,,
\end{equation}
which allows one to rewrite sums of diagrams with multiple emissions as products of diagrams with a single one. In non-abelian gauge theories such as QCD, there are genuine higher-order corrections to soft matrix elements since the different diagrams, and therefore the different terms on the left-hand side of \eqref{eik}, have different color structures and cannot be combined. However, the higher-order corrections only involve certain maximally non-abelian color structures \cite{Gatheral:1983cz,Frenkel:1984pz,Mitov:2010rp,Gardi:2010rn,Gardi:2013ita}.

While the inclusive cross section is finite, the hard and soft functions in \eqref{fact} individually suffer from divergences. The soft function suffers from ultraviolet (UV) divergences, which can be regularized using dimensional regularization. These UV divergences can be absorbed into the Wilson coefficients of the effective theory, which are encoded in the hard function. This renormalization renders the hard function finite, at the expense of introducing a renormalization scale $\mu$, which in our context is often called the factorization scale. After renormalization the theorem takes the form
\begin{equation}\label{factRen}
\sigma = \mathcal{H}(m_e, \{ \underline{v} \} ,\mu)\, \mathcal{S}(E_s,  \{ \underline{v} \}, \mu)\,,
\end{equation}
and the $\mu$ dependence of the functions fulfills an RG equation. Since the cross section is finite, the hard and soft anomalous dimensions must be equal and opposite. On a more concrete level, one observes that the on-shell amplitudes which define the hard function suffer from infrared (IR) divergences which cancel against the UV divergences of the soft function. Since the soft function exponentiates, also the divergences in the hard amplitudes must have this property.

\section{Expansion of loop integrals and the method of regions}
\label{region}

{\footnotesize
\begin{figure}[t!]        
\begin{center}              
\begin{overpic}[scale=0.8]{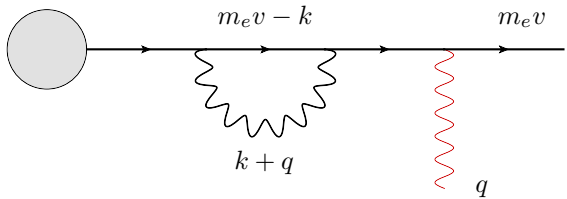}
\put(38,30){$m_e v-k$}
\put(41,4){$k+q$}
\put(88,30){$m_e v$}
\put(84,0){$q$}
\end{overpic}\\[10pt]
\end{center}
\vspace{-0.4cm}
\caption{Loop correction on an a fermion line.}   
\label{loopel}  
\end{figure}

}

When constructing $\mathcal{L}_{\rm eff}$, we have expanded in the soft photon momenta. This is fine for tree-level diagrams, but how about loops? Of course, the Taylor expansion does {\em not} commute with the loop integrations and to correct for this, one has to perform matching computations. We will now see that the part which gets lost in the naive low-energy expansion can be obtained by expanding the loop integrand in the region of large loop momentum. This is an example of a general technique called the {\em method of regions} \cite{Beneke:1997zp,Smirnov:2002pj} to expand loop integrals around various limits. We will use this method when constructing SCET, but it is instructive to discuss it  with a simple example integral in QED.

Let us consider the diagram shown in Fig.\ \ref{loopel}. For our discussion the numerators of the diagrams will not lead to complications and for simplicity we will therefore consider the associated scalar integral
\begin{equation}
F = \int d^dk \frac{1}{(k+q)^2} \frac{1}{(m_e v-k)^2-m_e^2 }\,,
\end{equation}
where $q$ is a soft-photon momentum. In the low-energy theory, we assumed that $k^\mu \sim q^\mu \ll m_e$. Expanding the integrand in this region gives integrals of the form
\begin{equation}\label{Flow}
F_{\rm low} = \int d^dk \frac{1}{(k+q)^2} \frac{1}{-2 m_e v\cdot k} \left\{ 1+ \frac{k^2}{2m_e v\cdot k} + \dots \right\}   \,,
\end{equation}
The expansion produces exactly the linear propagators $i/{v\cdot k}$ encountered in our tree-level discussion in the previous section. Of course, the loop momentum is integrated all the way to infinity and the expansion we have performed is no longer valid when $k^\mu \sim m_e$ or larger. One could follow Wilson and work with a hard cutoff to ensure that the loop momentum would never be too large, but we want to use dimensional regularization for the low-energy theory and not restrict the loop momentum. Looking at the integrals \eqref{Flow} we see that the expansion has produced ultraviolet divergences which are stronger than the one in the original integral, but the integrals are well-defined in dimensional regularization. 

To correct the problems from naively expanding the integrand, we consider the difference
\begin{align}
F_{\rm high} &\equiv F - F_{\rm low}   \\
&= \int d^dk \frac{1}{(k+q)^2}\left[ \frac{1}{(m_e v-k)^2-m_e^2 } - \frac{1}{-2 m_e v\cdot k} \left\{ 1+ \frac{k^2}{2m v\cdot k} + \dots \right\}  \right] \,.  \nonumber
\end{align}
By construction, the integrand has only support for $k^\mu \gg q^\mu$ since the square bracket tends to zero for $k^\mu\sim q^\mu$. We can therefore expand the integrand around $q^\mu = 0$ by expanding the first propagator denominator
\begin{align}
F_{\rm high} = \int d^dk \frac{1}{k^2} &\left\{ 1 - \frac{q^2}{2 q\cdot k} + \dots \right\}\, \nonumber\\
& \hspace{0cm}\, \times \left[ \frac{1}{(m_e v-k)^2-m_e^2 } - \frac{1}{-2 m_e v\cdot k} \left\{ 1+ \frac{k^2}{2m v\cdot k} + \dots \right\}  \right] \,.  \nonumber
\end{align}
Now we can evaluate the integrals one-by-one to get the high-energy part. What simplifies this task is that integrals of the form
\begin{equation}
I(\alpha,\beta,\gamma) = \int d^dk \, \left(k^2\right)^\alpha \, \left(v\cdot k\right)^\beta \left(q\cdot k\right)^\gamma = 0
\end{equation}
all vanish. To show this, rescale $k \to \lambda\, k$ . This yields $I(\alpha,\beta,\gamma)= \lambda^{d+2\alpha+\beta+\gamma}\, I(\alpha,\beta,\gamma)$ for any $\lambda >0$. Dropping the scaleless integrals, we get
\begin{align}
F_{\rm high} = \int d^dk \frac{1}{k^2} &\left\{ 1 - \frac{q^2}{2 q\cdot k} + \dots \right\}\, \frac{1}{(m_e v-k)^2-m_e^2 } \,, 
\end{align}
which is the expansion of the integrand for $k^\mu \sim m_e \gg q^\mu$. So we observe that we obtain the full result by performing the expansion of the integrand in two regions (low and high $k^\mu$), integrating each term and adding the two results. 

We can summarize the method of regions expansion as follows:
\begin{enumerate}[label=(\alph*),leftmargin=3\parindent]
\item Consider all relevant scalings (``regions'') of the loop momenta. In our example the scalings are $k_\mu \sim m_h$ (``hard region'') and $k_\mu \sim q_\mu$  (``soft region'').
\item Expand the loop integral in each region.
\item Integrate each term over the full phase space $\int \!d^dk$.
\item Add up the contributions.
\end{enumerate}
This technique provides a general method to expand loop integrals around different limits and can be used in many different kinematical situations \cite{Smirnov:2002pj}. Seeing that one integrates every region over the full phase space, one could be worried that this would lead to a double counting, but this is not the case. As our construction shows, the overlap region is given by scaleless integrals which can be dropped as we did in the last step. For this to be true, it was important that we consistently expanded away small momenta in the low energy region. Because of this, we ended up with single scale integrals, which become scaleless upon further expansion. If this is not done, one will need to eliminate the overlap region using subtractions, which are also called zero-bin subtractions in the context of SCET \cite{Manohar:2006nz}. A second important ingredient for the method is that one has to ensure that the expanded integrals are properly regularized and in some cases dimensional regularization alone is not sufficient. The reader interested to learn more about these issues can consult \cite{Smirnov:2002pj,Jantzen:2011nz}.

The method of region technique has a close connection to effective field theories in that the low-energy regions correspond to degrees of freedom in the effective field theory and the expanded full theory diagrams are equivalent to effective-theory diagrams, as we have seen in the example of soft effective theory. The contribution from the hard region gets absorbed into the Wilson coefficients. In the next section, we will apply the method of regions technique to the Sudakov form factor integral, which will allow us to identify the degrees of freedom relevant in this case. In addition to the soft region, we will find contributions from momentum configurations where the loop momentum is collinear to external momentum. As a consequence, the relevant effective theory then contains not only soft but also collinear particles.

\chapter{Soft-collinear effective theory}

In this chapter, we will go over the construction of the effective theory in detail. To do so, we will consider the simplest problem in which both soft and collinear particles play a role, namely the Sudakov form factor. By itself this is not a physical quantity, but it arises as a crucial element in many collider processes.

\section{Method of regions for the Sudakov form factor}
\label{RegionsSudakov}

Figure \ref{sudakov} shows the one-loop contribution to the Sudakov form factor. We define $L^2 = -l^2 - i0$, $P^2 = -p^2 - i0$ and $Q^2 = -(l-p)^2 - i0$ and will analyze the form factor in the limit
\begin{equation}\label{limit}
L^2 \sim P^2 \ll Q^2\,.
\end{equation}
This is the limit of large momentum transfer and small invariant mass, the same kinematics which is relevant for the jet process depicted in Fig.\ \ref{twojet}. Indeed, the corresponding loop correction will also arise in the computation of the jet cross section. The small off-shellness of the external lines arises in this case because of soft and collinear emissions from these lines. If the quantities $Q^2$, $P^2$ and $L^2$ are all positive, the form factor is real and analytic, the results in other regions can be obtained by analytic continuation, taking into account the $i0$ prescriptions specified above \eqref{limit}.

We want to find out which momentum modes are relevant in the Sudakov problem and how the different components of the momenta scale compared to the external momenta. Tensor loop integrals involve exactly the same momentum regions as scalar ones since all the complications arise from expanding propagator denominators. For our purposes it is therefore sufficient to study the scalar loop integral
\begin{equation}\label{fullint}
I  =   i \pi^{-d/2} \mu^{4-d} \int d^dk \frac{1}{\left(k^2 + i0\right) \left[(k+p)^2 + i0\right] \left[(k+l)^2 + i0\right]}\,.
\end{equation}
For convenience, we have included a prefactor to make the integral dimensionless; in the computation of the full  diagram the associated scale $\mu$ arises from the coupling in $d=4-2\ep$ dimensions.

{\footnotesize
\begin{figure}[th]        
\begin{center}              
\begin{overpic}[scale=1.0]{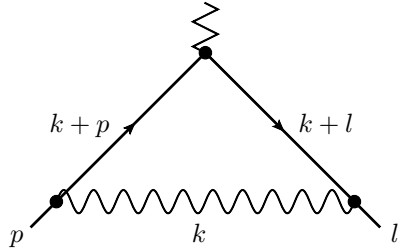}
\put(6,28){$k+p$}
\put(46,-3){$k$}
\put(76,28){$k+l$}
\put(102,-3){$l$}
\put(-5,-3){$p$}
\end{overpic}\\[10pt]
\end{center}
\caption{One-loop contribution to the Sudakov form factor.}   
\label{sudakov}    
\end{figure}
}

To perform the expansion of the integral around the limit \eqref{limit}, it is useful to introduce light-like reference vectors along $p^\mu$ and $l^\mu$, in analogy to the vectors $v_i^\mu$ we introduced in our discussion of soft photons. To be explicit, let's choose our coordinate system such that
\begin{align}
n^\mu & = ( 1,0,0,1) \approx p^\mu/p^0 \,,\\
\bar{n}^\mu & = ( 1,0,0,-1) \approx l^\mu/l^0\,,
\end{align}
with $n^2 = \bar{n}^2 =0$ and $n\cdot  \bar{n}  = 2$. Using these light-cone vectors any four vector can be decomposed in the form\begin{align} \label{momcomp}
p^\mu &= n\cdot p \frac{\bar{n}^\mu}{2} +\bar{n}\cdot p \frac{n^\mu}{2} +p^\mu_\perp \nonumber \\
 &=\;\;\; p_+^\mu \;\;\;\; +\;\;\;\; p_-^\mu\;\; \;\;+ p^\mu_\perp\,.
\end{align}
The quantities $(n\cdot p, \bar{n}\cdot p, p_\perp^\mu)$ are the light-cone components of the vector and we will now discuss how these components scale. To do so, it is useful to define a small expansion parameter
\begin{equation}
\lambda^2 \sim  P^2/Q^2  \sim L^2/Q^2 \ll 1\,.
\end{equation}
Note that
\begin{equation}
p^2 = n\cdot p  \, \bar{n}\cdot p + p_\perp^2\,.
\end{equation}
and due to $ p^2 \sim \lambda^2 Q^2 $ and $p^\mu = p_-^\mu + \mathcal{O}(\lambda) = Q/2\, n^\nu  + \mathcal{O}(\lambda)$ the components of the external momenta must scale as
\begin{align}
 		& (\,n\cdot p, \bar{n}\cdot p, \; p_\perp^\mu) \nonumber \\
p^\mu \;\sim\; \;& (\,\;\; \lambda^2 \,,\; \;1\; \;\;,\;\; \lambda \;) \;Q \,,\nonumber \\
l^\mu \;\sim \; \;& (\,\;\; 1 \;\;, \;\; \lambda^2\;\;  , \;\; \lambda \;) \; Q\,.
\end{align}
In our context the term {\em scaling as $\lambda^a$} means that the given quantity approaches the limit $\lambda \to 0$ as the $a$-th power of the expansion parameter $\lambda$.

We now perform the region expansion of our integral after assigning different scalings to the loop momentum $k^\mu$ and expanding the integrand in each region, proceeding exactly as outlined at the end of Section \ref{region}. The following scalings yield nonzero contributions
\begin{equation}\label{SCETregions}
\begin{aligned}
  &	&	 & (n\cdot k, \bar{n}\cdot k, k_\perp^\mu\,)  \\
&\text{hard} &(h)\quad &   (\,\;\; 1\;\; \;,\; \;1\; \;\;,\;\; 1 \;) \;Q \,, \\
&\text{collinear to $p^\mu$} &(c)\quad &  (\,\;\; \lambda^2 \;,\; \;1\; \;\;,\;\; \lambda \,) \; Q\,,  \\
&\text{collinear to $l^\mu$} &(\bar{c})\quad &  (\,\;\; 1 \;\;\,, \;\; \lambda^2\;\;  , \; \lambda\, \;) \; Q \,, \\
&\text{soft} &(s)\quad &  (\,\; \lambda^2 \;\,, \;\; \lambda^2\;\;  , \; \lambda^2 ) \; Q\,.
\end{aligned}
\end{equation}
For brevity, we will refer to the momenta which are collinear to $p^\mu$ simply as {\em collinear} and to the ones collinear to $l^\mu$ {\em anti-collinear}. All other scaling choices $(\lambda^a , \lambda^b, \lambda^c)$ for the loop momentum lead to scaleless integrals upon performing the expansion -- pick one and check! 

Since in the soft region all components of the loop momentum scale as $\lambda^2$, we have
\begin{equation}
k_s^2 \sim \lambda^4 Q^2 \sim \frac{P^2 L^2}{Q^2}  \ll P^2 \sim L^2 \,.
\end{equation}  
This is the hierarchy we advertised in \eqref{scales}, and since $k_s^2 \ll k_c^2$, this mode is sometimes also called ultra-soft. For some other observables, the soft mode scales as $(\lambda , \lambda, \lambda)$. The version of SCET for this situation is called SCET$_{\rm II}$ to distinguish it from the one relevant for the Sudakov form factor which is also called SCET$_{\rm I}$. SCET$_{\rm II}$ involves so-called rapidity logarithms which are not present in SCET$_{\rm I}$ and there are different formalisms to deal with them. Direct exponentiation based on the {\em collinear anomaly} \cite{Becher:2010tm}, or resummation using the {\em rapidity renormalization group} \cite{Chiu:2011qc,Chiu:2012ir}. For our introduction, we will exclusively work with SCET$_{\rm I}$. The reader interested to learn more about SCET$_{\rm II}$ can consult the book \cite{Becher:2014oda}. 

Let us now expand the integrand in the different regions to leading power. To get the leading-power integrand in the hard region, we simply set all the suppressed momentum components to zero, which amounts to replacing $p^\mu \to p_-^\mu$ and $l^\mu \to l_+^\mu$. This leads to
\be \label{eq:Ihard}
I_h = i \pi^{-d/2} \mu^{4-d} \int d^d k \frac{1}{\left(k^2 +i0\right)
\left(k^2 +2 k_-\cdot l_+ +i 0 \right) \left( k^2 + 2 k_+\cdot p_- +i 0 \right)} \, .
\ee
The corresponding loop integral is standard; it is simply the on-shell form factor integral. Performing it in the usual way, one obtains
\bea\label{eq:Ihardres}
I_h &=& \frac{\Gamma(1+\varepsilon)}{2 l_+ \cdot p_-} \frac{\Gamma^2(-\varepsilon)}{\Gamma(1 -2 \varepsilon)} 
\left(\frac{\mu^2}{2  l_+ \cdot p_-} \right)^\varepsilon \,  .
\eea
For  $\ep\to 0$, one encounters soft and collinear infrared divergences when evaluating the integral. 

In the collinear region the integration momentum scales as 
$k^\mu \sim (\lambda^2, 1, \lambda)\, Q$ and $k^2 \sim \lambda^2 Q^2$. We can therefore expand 
\be
(k+l)^2 = 2 k_-\cdot l_+ +{\mathcal O}(\lambda^2)\,,
\ee
while the other two propagators stay as they are. We end up with the integral
\bea \label{eq:Icp}
I_c &=&  i \pi^{-d/2} \mu^{4-d} \int d^d k \frac{1}{\left(k^2 +i0\right)
\left(2 k_-\cdot l_+ +i 0 \right) \left[ (k+p)^2 +i 0 \right]} \, .
\eea
This loop integral involves a propagator which is linear in the loop momentum. To deal with it, one can use a variant of the usual Feynman parametrization, 
\begin{equation}\label{feynp}
 \frac{1}{A^n B^m} = \frac{\Gamma (m+n)}{\Gamma (m) \Gamma (n)} \int_0^\infty d\eta \frac{\eta^{m-1}}{(A+\eta B)^{n+m}} \,,
\end{equation}
where $B$ is the linear propagator, while $A$ is quadratic. Using this parametrization, one again ends up with a standard loop integral. Evaluating it yields
\bea \label{Icoll}
I_c &=& -\frac{\Gamma(1+\varepsilon)}{2 l_+ \cdot p_-} \frac{\Gamma^2(-\varepsilon)}{\Gamma(1- 2 \varepsilon)} \left( \frac{\mu^2}{P^2} \right)^\varepsilon \, .
\eea
While it is some work to evaluate the integral, at least its scaling is easily obtained. $I_c$ can depend on the invariants $P^2$ and $ l_+ \cdot p_-$. Looking at the integrand one observes that  $I_c \to I_c /\alpha $ under a rescaling $ l_+ \to  \alpha\, l_+$. Together with dimensional analysis, this dictates the dependence on $P^2$ so that the only unknown is the $\varepsilon$-dependent prefactor. The contribution of the $\bar{c}$-region is obtained by replacing $P^2 \to L^2$ in \eqref{Icoll}. This leaves the soft region in which $k^\mu \sim (\lambda^2, \lambda^2 , \lambda^2)\, Q$ so that we can approximate
\begin{align}
 (k+l)^2 &= 2 k_-\cdot l_+ + l^2 + {\mathcal O}(\lambda^3)\, , &  \quad(k+p)^2 &= 2 k_+\cdot p_- + p^2 + {\mathcal O}(\lambda^3)\, .
\end{align}
Dropping the higher order terms, we are now left with two linear propagators, which can be handled by using \eqref{feynp} twice. Performing the loop integration and integrating over the Feynman parameters one obtains
\bea \label{eq:Isres}
I_{s} &=& i \pi^{-d/2} \mu^{4-d} \int d^d k \frac{1}{\left(k^2 +i0\right)
\left(2 k_-\cdot l_+ +l^2 +i 0 \right) \left(2 k_+\cdot p_- +p^2  +i 0 \right)} \,  \nn \\
&=& - \frac{\Gamma\left(1+\varepsilon\right)}{2 l_+ \cdot p_-} \Gamma(\varepsilon) \Gamma\left( - \varepsilon\right) \left( \frac{2  l_+\cdot p_- \mu^2}{L^2 P^2}\right)^\varepsilon \,  .
\eea

Having obtained the contributions from all the different momentum regions, we can now add them up and verify whether we reproduce the full integral \eqref{fullint}. This involves some nontrivial cancellations since the individual integrals are all divergent, while the full integral is finite in $d=4$. To make the divergences explicit, let us expand in $\varepsilon$ and list the individual contributions. 
\bea
I_{h} &=& \frac{\Gamma\left(1+\varepsilon\right)}{Q^2}
\left(\frac{1}{\varepsilon^2} +\frac{1}{\varepsilon} \ln\frac{\mu^2}{Q^2}
+\frac{1}{2} \ln^2 \frac{\mu^2}{Q^2} -\frac{\pi^2}{6} \right) \nn \\
I_{\cc} &=& \frac{\Gamma\left(1+\varepsilon\right)}{Q^2}
\left(-\frac{1}{\varepsilon^2} -\frac{1}{\varepsilon} \ln\frac{\mu^2}{P^2}
-\frac{1}{2} \ln^2 \frac{\mu^2}{P^2} +\frac{\pi^2}{6} \right)\nn \\
I_{\cb} &=& \frac{\Gamma\left(1+\varepsilon\right)}{Q^2}
\left(-\frac{1}{\varepsilon^2} -\frac{1}{\varepsilon} \ln\frac{\mu^2}{L^2}
-\frac{1}{2} \ln^2 \frac{\mu^2}{L^2} +\frac{\pi^2}{6} \right) \nn \\
I_{s} &=& \frac{\Gamma\left(1+\varepsilon\right)}{Q^2}
\left(\frac{1}{\varepsilon^2} +\frac{1}{\varepsilon} \ln\frac{\mu^2 Q^2}{L^2 P^2}
+\frac{1}{2} \ln^2 \frac{\mu^2 Q^2}{L^2 P^2} +\frac{\pi^2}{6} \right) \nn \\ 
&&\hspace{-0.8cm}\makebox[9.5cm]{\hrulefill}\nn \\[3pt]
I_{\rm tot} &=&  \frac{1}{Q^2}
\left(\ln{\frac{Q^2}{L^2}} \ln{\frac{Q^2}{P^2}}+ \frac{\pi^2}{3} \right) \, .
\eea
The sum $I_{\rm tot}  =   I_h + I_{\cc} + I_{\cb} + I_s$ is indeed finite and reproduces the leading term in the expansion of the full integral $I$. The cancellation of divergences is especially remarkable for the $1/\varepsilon$ pieces which all involve logarithms of different scales which must add up to zero. The cancellation  implies that the infrared divergences of the hard integrals are the same as the ultra-violet divergences of the SCET matrix elements. Similar relations arise for $n$-point Green's functions and imply strong constraints on the infrared divergences of on-shell amplitudes in QCD and other gauge theories. These were analyzed in \cite{Becher:2009cu,Becher:2009qa,Becher:2009kw,Ahrens:2012qz} using SCET and with diagrammatic methods in \cite{Gardi:2009qi,Dixon:2009ur,Bret:2011xm,DelDuca:2011ae}. The constraints fully fix the infrared structure of an arbitrary two-loop $n$-point amplitudes in terms of quantities which can be extracted from Sudakov form factors. A new structure, which involves four legs, first arises at three loops and was recently computed \cite{Almelid:2015jia,Henn:2016jdu,Almelid:2017qju}.

\section{Effective Lagrangian}

Using the form factor integral, we have identified the relevant momentum regions for the Sudakov problem. We will now proceed similar to Chapter \ref{set} and construct a Lagrangian whose Feynman rules directly yield the expanded diagrams obtained using method of regions expansion. To do so, we introduce effective field theory fields $\phi_c$, $\phi_{\bar{c}}$ and $\phi_s$ (where $\phi$ denotes a quark or gluon field)  whose momenta scale exactly as appropriate for the relevant momentum region. 

The standard approach to constructing an effective theory is to write down the most general Lagrangian for the given fields and to then adjust the Wilson coefficients to reproduce the low-energy behavior of the full theory. At tree level, we can use a short-cut and simply substitute
\begin{equation}
\begin{aligned}\label{treesub}
\psi & \to \psi_c + \psi_{\bar{c}} + \psi_s \,, \\
A^\mu & \to A^\mu_c  + A^\mu_{\bar{c}} + A^\mu_s\,,
\end{aligned}
\end{equation}
in QCD, expand away the suppressed terms and read off the effective Lagrangian. At loop level, matching corrections arise, which can modify the coefficients of the tree-level operators and induce new operators, but we will find that the matching corrections only affect the terms involving both $c$ and $\bar{c}$ fields. To construct the purely soft and collinear Lagrangians, the tree-level short cut \eqref{treesub} is useful and efficient.  

Both the fermion and gauge fields have several components, which scale differently with the expansion parameter $\lambda$. To read off the scaling of the fields, it is easiest to consider the propagators. Let us start with the gluon-field propagator
\begin{equation}
\langle 0|\, T\!\left\{ A_\mu^a (x) A_\nu^b (0) \right\} |0 \rangle =\int \frac{d^4 p}{(2 \pi)^4} \frac{i}{p^2 +i 0} e^{-i p\cdot x} \left[ - g_{\mu \nu} +\xi \frac{p_\mu p_\nu}{p^2}\right] \delta^{ab}\,.
\end{equation}
Here $a$ and $b$ are the color indices of the fields. In the following, we will usually work with matrix fields $A_\mu (x)=A_\mu^a (x) t^a$, where the $t^a$'s are the generators of the gauge group, and keep the color implicit. The position argument $x^\mu$ of the fields is conjugate to the momentum $p^\mu$ in the Fourier exponent $p\cdot x \sim \mathcal{O}(\lambda^0)$. The part of the propagator involving the gauge parameter $\xi$ scales like $d^4p /(p^2)^2\, p_\mu p_\nu \sim p_\mu p_\nu $. In a generic gauge, the gluon field thus scales exactly like the momentum $A_\mu \sim p_\mu$. This is of course expected since gauge symmetry ties the field to the momentum. For soft and collinear gluons, the field component thus scale as
\begin{equation}
\begin{aligned}
\left (n\cdot  A_s\,, \, \bar{n}\cdot  A_s\,, \,A_{s\perp}^\mu \right) &\sim \left( \lambda^2\,,\, \lambda^2\,,\, \lambda^2\right),\\
\left (n\cdot  A_c\,, \, \bar{n}\cdot  A_c\,,\, A_{c\perp}^\mu \right) & \sim  \left(  \lambda^2\,,\, 1\;\,,\; \lambda\;\, \right).
\end{aligned}
\end{equation}
From the scaling we immediately see that in terms involving both soft and collinear gluons, the soft gluons are power suppressed, except for the contribution from the $n\cdot  A_s$ component, which is commensurate with its collinear counterpart. 

Next, let's consider the fermion propagators. For a soft fermion, we have
\begin{equation}
\langle 0|\, T\!\left\{ \psi_s (x) \bar{\psi}_s (0) \right\} |0 \rangle = \int \frac{d^4 p}{(2 \pi)^4} \frac{ip\!\!\!\slash}{p^2 +i 0} e^{-i p\cdot x}  \sim (\lambda^2)^4 \frac{ \lambda^2}{\lambda^4} = \lambda^6 \, ,
\end{equation}
from which we conclude that the soft fermions scale as $ \psi_s(x)\sim \lambda^3$. The collinear case is more complicated because the numerator of the propagator must be decomposed into
\begin{equation}
p\!\!\!/ = n\cdot p \frac{\bar{n}\!\!\!/}{2} +\bar{n}\cdot p \frac{n\!\!\!/}{2} +{p\!\!\!/}^\perp
\end{equation}
and the three terms have different scaling, which implies that different parts of the fermion spinor scale differently. To take this into account, one splits the fermion field into two parts
\begin{equation}
\psi_c = \xi_c + \eta_c =  P_+ \,\psi_c + P_- \,\psi_c
\end{equation}
using the projection operators
\begin{align}
P_+ &= \frac{\nsl \nbsl}{4} \,,& P_- &= \frac{ \nbsl \nsl}{4}\,,
\end{align}
which fulfill $P_+ + P_- = 1$ and $P_\pm^2 = P_\pm$. The propagator
\begin{equation}
\begin{aligned}
\langle 0| T\left\{ \xi_c (x) \bar{\xi}_c (0) \right\} |0 \rangle &= \int \frac{d^4 p}{(2 \pi)^4} e^{-i p\cdot x} \frac{\nsl \nbsl}{4} \frac{i p\!\!\!\slash}{p^2 +i 0} \frac{ \nbsl \nsl}{4} \\
&= \int \frac{d^4 p}{(2 \pi)^4} e^{-i p\cdot x} \frac{i \bar{n}\cdot p \frac{n\!\!\!/}{2}}{p^2 +i 0}   \sim \lambda^4\frac{1}{\lambda^2} = \lambda^2 \, ,
\end{aligned}
\end{equation}
shows that the field scales as $ \xi_c \sim \lambda$ and repeating the exercise for $\eta_c$ one reads off that $\eta_c\sim \lambda^2$. We observe that soft fermions are power suppressed with respect to collinear fermions.

Now that we know the scaling of all fields, we can plug the decomposition \eqref{treesub} into the QCD action and read off the tree-level effective field theory Lagrangian, 
\begin{equation}
S = S_s + S_c + S_{\bar{c}} + S_{c+s} + S_{\bar{c} +s}  + \dots
\end{equation}
where we collected the terms according to their field content. $S_s$ contains the purely soft terms, $S_c$  the collinear terms, and $S_{c +s}$ describes the soft-collinear interactions. Since the $S_{\bar{c}}$ and $S_{\bar{c}+s}$ can be obtained from the collinear terms using simple substitution rules, we will suppress them for the moment.

Let us first write the purely soft part, which has exactly the same form as the standard QCD action, except that the field is replaced with the soft field
\begin{equation}\label{Ssoft}
S_s = \int d^4x \, \bar{\psi}_s i \Dsl_s \psi_s -\frac{1}{4}(F^{a}_s)_{\mu \nu} (F^{a}_s)^{\mu \nu}\,,
\end{equation}
where the soft covariant derivative is $i D^\mu_s = i\partial^\mu + A_s^\mu$ and the soft field strength tensor $(F^{a}_s)_{\mu \nu}$ is of course built from this derivative. Note that all the terms in the action are $\mathcal{O}(\lambda^0)$ since the integration measure $d^4x$ counts as $\mathcal{O}(\lambda^{-8})$ because the position vector components scale as $x^\mu \sim 1/ p_s^\mu \sim \lambda^{-2}$. That we reproduce the standard QCD action is of course expected. The purely soft quarks and gluons behave exactly as standard quarks and gluons and since everything scales in the same way, there is nothing which can be expanded away. Next, let's consider the purely collinear part of the action. This is again a simple copy of the QCD action, but we write out the fermion field in its two components $\psi_c = \xi_c + \eta_c$ and get
\begin{equation}\label{Scoll}
\begin{aligned}
S_c &=\int d^4x \left(\bar{\xi}_c + \bar{\eta}_c\right)
\left[\frac{\nsl}{2} i \bar{n} \cdot D_c  + \frac{\nbsl}{2} i n \cdot D_c +
i \Dsl_{c\perp} \right] \left(\xi_c + \eta_c \right)   -\frac{1}{4}(F^{a}_c)_{\mu \nu} (F^{a}_c)^{\mu \nu} \,  \\
&=\int d^4x  \, \bar{\xi}_c \,\frac{\nbsl}{2} i n \cdot D_c\, \xi_c + 
\bar{\xi}_c\, i \Dsl_{c\perp}\, \eta_c + \bar{\eta}_c \, i\Dsl_{c\perp}\, \xi_c +
\bar{\eta}_c \,\frac{\nsl}{2} i \bar{n} \cdot D_c \, \eta_c   -\frac{1}{4}(F^{a}_c)_{\mu \nu} (F^{a}_c)^{\mu \nu}\,.
\end{aligned}
\end{equation}
Using the scaling of the fields and taking into account that the integration measure now scales as $d^4x \sim \lambda^{-4}$ because the position is conjugate to a collinear momentum, the reader easily verifies that each term is $\mathcal{O}(\lambda^0)$. This form of the Lagrangian is inconvenient since it involves both the large component $\xi_c \sim \lambda$ and the power suppressed field $\eta_c \sim \lambda^2$ and the two mix into each other. To construct operators, it is simplest to avoid this complication by integrating out the small components of the fermion field, similar to what is done when constructing the Lagrangian in Heavy-Quark Effective Theory (HQET). In HQET, the small component has a propagator which scales as $1/m_q$, where $m_q$ is the heavy quark mass. In this sense, the small component is not really a dynamical field and can be removed. In close analogy, the propagator of the small component of the collinear field scales as $1/Q$ and can therefore be integrated out. In practical computations, it can be convenient to keep the small component since one can then use the standard QCD Feynman rules for collinear computations, as was done e.g.\ in \cite{Becher:2006qw}, and some authors have advocated to formulate the theory \cite{Freedman:2011kj} without integrating out the small components. At any rate, the action \eqref{Scoll} is quadratic in $\eta_c$ and we can therefore integrate out the field exactly. To do so, one shifts the field
\begin{equation}\label{shiftColl}
\eta_c \to \eta_c - \frac{\nbsl}{2} \frac{1}{i \bar{n}\cdot D_c}  i\Dsl_{c\perp}\, \xi_c
\end{equation}
to complete the square in the action $S_c$. After the shift, the action takes the form
\begin{equation}\label{Lcoll}
\mathcal{L}_c = \bar{\xi}_c \frac{\nbsl}{2} \left[ i n \cdot D_c +   i\Dsl_{c\perp} \frac{1}{i \bar{n}\cdot D}  i\Dsl_{c\perp} \right] \xi_c   -\frac{1}{4}(F^{a}_c)_{\mu \nu} (F^{a}_c)^{\mu \nu} + \bar{\eta}_c \,\frac{\nsl}{2} i \bar{n} \cdot D_c \, \eta_c\,
\end{equation} 
and we can integrate out the field $\eta_c$, which leaves a determinant $\det (\frac{\nsl}{2} i \bar{n} \cdot D_c)$. To make the determinant and the inverse derivative in the shift \eqref{shiftColl} well defined, we should adopt an $i0$ prescription for the $i \bar{n}\cdot D_c$ operator. The prescription is without physical consequences since it concerns the region near $\bar{n}\cdot p =0$, while the effective theory deals with processes with $\bar{n}\cdot p \sim Q$. It turns out that the determinant is trivial, which can be seen by evaluating it in the gauge $\bar{n} \cdot A_c=0$ and noting that it is gauge invariant, or by realizing that the associated closed-loop diagrams all vanish since for a given $i0$ prescription all propagator poles in the closed $\eta_c$ loops are on the same side of the integration contour and can be avoided. After dropping the trivial determinant, the collinear SCET Lagrangian is therefore given by \eqref{Lcoll} without the last term. 

Next we consider the soft-collinear interaction terms in $S_{s+c}$. The general construction of the terms is quite involved and was performed in \cite{Beneke:2002ni} in the position-space formalism we are using here. For simplicity, we will restrict ourselves to leading-power terms which are usually sufficient for collider physics applications. Nevertheless there is currently a lot of effort to analyze power corrections, the interested reader can consult the recent papers \cite{Moult:2016fqy,Boughezal:2016zws,Bonocore:2016awd,Feige:2017zci,Moult:2017jsg,Goerke:2017lei,Beneke:2017ztn} for more information. Getting the leading-power soft-collinear interactions is quite simple once we remember the power counting of the fields discussed above
\begin{itemize}
\item $\psi_s$ is power suppressed compared to the collinear fermion fields, so that no soft quarks are present in leading-power interactions with collinear fields.
\item Since $\bar{n}\cdot  A_s$ and  $A_{s\perp}^\mu$ are power suppressed compared to their collinear counterparts, only the component  $n\cdot  A_s$ arises.
\end{itemize}
Taken together, this implies that the leading power interaction terms can be obtained by substituting
\begin{equation}\label{softsub}
A_c^\mu \to A_c^\mu + n\cdot  A_s \frac{\bar{n}^\mu}{2}
\end{equation}
in $S_c$. 

The final step is to perform a derivative expansion in the resulting Lagrangian which corresponds to the expansion in small momentum components. To do so, consider the interaction term of a soft gluon with a collinear fermion
\begin{equation}
S_{c+s} = \int\! d^4x \,\bar{\xi}_c(x) \frac{\nbsl}{2}   n \cdot A_s(x)\, \xi_c(x)\,,
\end{equation}
which arises from the substitution \eqref{softsub} into \eqref{Lcoll}, together with purely gluonic interactions. The momentum is given by a soft and a collinear momentum which scales like a collinear momentum so that $x^\mu$ is conjugate to a collinear momentum. Explicitly, this implies the scaling
\begin{align}
p_c^\mu + p_s^\mu \sim p_c^\mu  &\sim  \left(\,  \lambda^2\,,\, 1\;\;\; \;,\, \lambda\;\; \;\right)\,, \nn \\
x^\mu  & \sim  \left(   \;1\;\;, \lambda^{-2}\,,\, \lambda^{-1} \; \right)\,,
\end{align}
which follows from the fact $p_c \cdot x = 1/2\, n\cdot p_c \bar{n}\cdot x +  1/2\,\bar{n}\cdot p_c n\cdot x +p_c^\perp\cdot x^\perp   \sim  1$. When we therefore consider a product of a soft momentum with the position vector
\begin{align}
p_s \cdot x & =  \frac{1}{2}\underbrace{n\cdot p_s\, \bar{n}\cdot x}_{\mathcal{O}(1)} \,+\,\frac{1}{2} \underbrace{  \bar{n}\cdot p_s \bar{n}\cdot x}_{\mathcal{O}(\lambda^2)}\, +\, \underbrace{ p_s^\perp\cdot   x^\perp}_{\mathcal{O}(\lambda)} \nn \\
&= \;\;\;\;p_{s+} \cdot x_-\;\; \;+\;\; p_{s-} \cdot x_+ \;\;\;\;+\;\;  p_\perp \cdot x_\perp
\end{align}
only the term involving $x_-$ is of leading power. We can therefore expand the interaction term into a Taylor series,
\begin{align}
S_{c+s} &= \int\! d^4x \,\bar{\xi}_c(x) \frac{\nbsl}{2} \xi_c(x)\,\Big [ 1 + \underbrace{x_\perp \cdot \partial_\perp}_{\mathcal{O}(\lambda)} +  \underbrace{x_+ \cdot \partial_-}_{\mathcal{O}(\lambda^2)}+ \dots \Big]   n \cdot A_s(x) \Big|_{x=x_-} \nn \\
&=   \int\! d^4x \,\bar{\xi}_c(x) \frac{\nbsl}{2} \xi_c(x) n \cdot A_s(x_-)  + \mathcal{O}(\lambda)\,,
\end{align}
and the leading power interactions are obtained by replacing $x^\mu \to x_-^\mu$ in the argument of the soft fields. This derivative expansion was called the {\em multipole expansion} in \cite{Beneke:2002ph}, since it has  similarity to what is done when approximating charge distributions at large distances in electrodynamics. Let us note that the original SCET papers \cite{Bauer:2000yr,Bauer:2001yt} and a large fraction of the current SCET literature uses a different method to expand in small momenta. In this approach, one splits the momenta into large and small components and treats the large components in Fourier space, while the small ones remain in position space. This is similar to the procedure we used in soft effective theory, where we split the electron momentum into $p^\mu = m_e v^\mu +k^\mu $, where $k^\mu$ is the soft residual momentum in which we expand. The position dependence of the field $h_v(x)$ is conjugate to the residual momentum $k^\mu$ and the large part $m_e v^\mu$ became a label on the field $h_v(x)$. For this reason, this hybrid momentum-and-position-space formulation of SCET is called the {\em label formalism}. We will work in position space and not cover the label formulation in more detail. The book \cite{Becher:2014oda} contains a comparison of the two formulations. At leading power, it is simple to translate between them.

After performing the multipole expansion, we arrive at the final form of the leading-power SCET Lagrangian
\be \label{eq:LSCET}
\mathcal{L}_{{\tiny \mbox{SCET}}} =\bar{\psi}_s i \Dsl_s \psi_s + 
\bar{\xi}_c \frac{\nbsl}{2} \left[ i n \cdot D + i \Dsl_{c \perp} \frac{1}{i \bar{n} \cdot D_c} i \Dsl_{c \perp}\right] \xi_c -\frac{1}{4} \left( F_{\mu \nu}^{s, a} \right)^2 -\frac{1}{4} \left( F_{\mu \nu}^{c, a} \right)^2 \, .
\ee
This expression involves the collinear and soft covariant derivatives
\begin{align} \label{eq:listcd}
i D^s_\mu &= i \partial_\mu + g A^{s }_\mu(x) \, ,&
i D^c_\mu &= i \partial_\mu + g A^{c }_\mu(x)  \, , 
\end{align}
as well as the mixed derivative
\begin{align}
i n\cdot D &= i n \cdot \partial + g\, n \cdot A_c (x) + g\,n \cdot A_s (x_-) \, .
\end{align}
The associated field-strength tensors are
\begin{align} \label{eq:fieldstr}
ig F^{s,a}_{\mu \nu} t^a &= \left[ i D_\mu^s ,  i D_\nu^s \right] \, ,&
ig F^{c,a}_{\mu \nu} t^a &= \left[ i D_\mu ,  i D_\nu \right] \, ,
\end{align}
where the derivative appearing in the second commutator in \eqref{eq:fieldstr} is defined as
\be \label{eq:DforF}
D^\mu = n \cdot D \frac{\bar{n}^\mu}{2} +\bar{n} \cdot D_c \frac{n^\mu}{2} + D^\mu_{c \perp} \, .
\ee
As stated above, we omitted the terms involving the $\bar{c}$ fields for brevity in the Lagrangian \eqref{eq:LSCET}. They have the same form as the ones involving $c$ fields, but with $n \leftrightarrow \bar{n}$ and $x_-  \leftrightarrow x_+$.

To finish the discussion of the Lagrangian, let us briefly discuss gauge transformations. Since we split the gauge field into different soft and collinear components, we can consider separate gauge transformations
\begin{align}
& \text{soft:} &	V_s(x) = \exp[ i \alpha_s^a(x) t^a]\,, \\
& \text{collinear:}  & V_c(x) = \exp[ i \alpha_c^a(x) t^a]\,,
\end{align}
where the gauge transformations scale like the associated fields $\partial_\mu \alpha_s^a(x) \sim \lambda^2 \alpha_s^a(x) $ and $\partial_\mu \alpha_c^a(x) \sim (\lambda^2,1, \lambda) \alpha_c^a(x)$. The soft gauge
transformations act on the soft fields in the usual way,\
\begin{equation}\label{gauge}
\begin{aligned}
\psi_s(x) & \to V_s(x) \,\psi_s(x)\,, \\
 A_s^\mu(x) &\to V_s(x) \, A_s^\mu(x)\, V^\dagger_s(x) + \frac{i}{g} V_s(x) (\partial_\mu V^\dagger_s(x) )\,,
\end{aligned}
\end{equation}
so that the covariant derivative transforms as $D_s^\mu(x)  \to V_s(x) \, D_s^\mu(x)\, V^\dagger_s(x)$ while the collinear fields transform as
\begin{equation}
\begin{aligned}
\psi_c(x) & \to V_s(x_-) \,\psi_c(x) \,, \\
 A_c^\mu(x) &\to V_s(x_-) \, A_c^\mu(x)\, V^\dagger_s(x_-) \,.
\end{aligned}
\end{equation}
This differs in two important aspects from the transformation of the soft fields. First of all, we have performed the multipole expansion and have replaced $x\to x_-$ in the soft fields. Without this expansion, the gauge transformations would induce a tower of power corrections, which would be inconvenient. Secondly, we observe that the collinear gauge field $A_c^\mu(x)$ transforms as a matter field, i.e.\ without the inhomogeneous term present in \eqref{gauge}. This ensures that the mixed derivative transforms in the correct way as
\begin{equation}
D^\mu \to V_s(x_-) \, D^\mu(x)\, V^\dagger_s(x_-)\,.
\end{equation}
Next, let's consider collinear transformations. Since these involve a collinear field which carries a large momentum, the soft fields must remain invariant under these transformations, while the collinear ones transform in the usual way
\begin{equation}
\begin{aligned}
\xi_c(x) & \to  V_c(x)  \,\xi_c(x)  &
D^\mu & \to V_c(x)\, D^\mu\, V^\dagger_c(x) \,,\\
\psi_s(x) & \to \psi_s(x) & D^\mu_s & \to  D^\mu_s \,.
\end{aligned}
\end{equation}
To see what kind of transformation this implies for the field $A_c^\mu(x)$, the reader can consult \cite{Becher:2014oda}. Given that the covariant derivatives transform in the expected way, it is straightforward to verify that the Lagrangian \eqref{eq:LSCET} is invariant under both soft and collinear gauge transformations.

Let us finally come back to the statement that there are no matching corrections to the Lagrangian. For the purely soft Lagrangian this is clear, since it is a copy of the QCD Lagrangian. All purely soft processes are therefore identically reproduced by the effective theory. For this to be true, it is important that we work with dimensional regularization rather than with a hard cutoff. If we would restrict the soft momenta to be below a certain cutoff in the loop diagrams of the low-energy theory, we would not reproduce the higher-energy parts of QCD and would need to correct for this by performing a matching computation. The same statements apply to the collinear Lagrangian. It looks different than the QCD Lagrangian because we integrated out two components, but since we did not perform any approximation, the collinear SCET Lagrangian is equivalent to QCD. One could worry that soft loops would contribute to purely collinear processes and that one would need to remove these contributions, but dimensional regularization is again very efficient in eliminating unnecessary contributions. It turns out that soft loop corrections to purely collinear diagrams are all scaleless and vanish. One can understand this by looking at the soft loop integral \eqref{eq:Isres}. It involves the scale $\Lambda_s^2 = P^2 L^2 /Q^2$, which is nonzero only when both $c$ and $\bar{c}$ particles are involved. Matching corrections are therefore only present for interactions involving both types of fields, such as the current operators discussed in the next section. 

\section{The vector current in SCET}

We have constructed $\mathcal{L}_s$, $\mathcal{L}_c$ and $\mathcal{L}_{s+c}$ in the previous section and can obtain $\mathcal{L}_{\bar{c}}$ and $\mathcal{L}_{s+\bar{c}}$ from simple substitutions in the $c$ terms. What is missing are operators involving both $c$ and $\bar{c}$ fields. In the Sudakov problem only the electromagnetic current operator connects the two fields. The necessary hard momentum transfer is provided by the virtual photon on the external line. The tree-level diagram in QCD can be reproduced by a SCET operator
\begin{equation}\label{treeCurrent}
\begin{aligned}
\put(88,-10){$p$}
\put(-10,-12){$l$}
\begin{overpic}[scale=0.7]{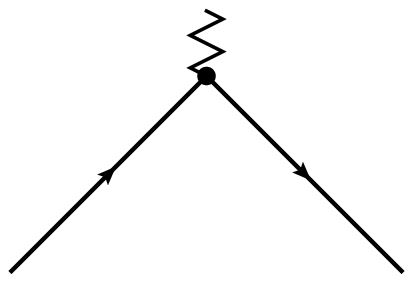}
\end{overpic}
\quad\quad
&\raisebox{0.7cm}{$\longrightarrow$}
\quad \quad
\begin{overpic}[scale=0.7]{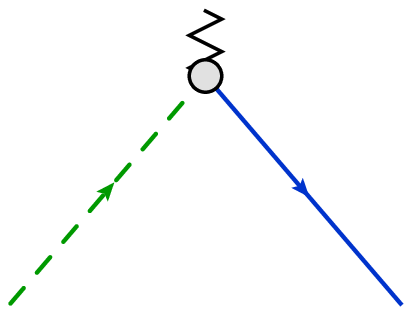}
\put(105,-10){$c$}
\put(-10,-12){$\bar{c}$}
\end{overpic} \\[12pt]
J^\mu = \bar{\psi}\, \gamma^\mu\, \psi \quad\quad\quad \quad
&\longrightarrow
\quad\quad\quad\quad  \bar{\xi}_c\, \gamma^\mu\, \xi_{\bar{c}} \quad\quad
\end{aligned}
\end{equation}
where we have indicated the $\bar{c}$ fermion with a green dashed line and the $c$ field with a blue line.
Due to the projection properties of the two fermion fields, we can simplify the SCET operator a bit further
\begin{equation}\label{opnaive}
\bar{\xi}_c\, \gamma^\mu\, \xi_{\bar{c}} = \bar{\xi}_c \left[n^\mu \frac{\bar{n}\!\!\!/}{2} +\bar{n}^\mu \frac{n\!\!\!/}{2} +{\gamma}_\perp^\mu  \right]  \xi_{\bar{c}} = \bar{\xi}_c\,{\gamma}_\perp^\mu \, \xi_{\bar{c}} \,.
\end{equation}
However, beyond tree-level and for processes involving collinear gluons, the above operator is insufficient and it will now take quite some work to write down the most general gauge-invariant leading-power operator. 

One problem with the operator in \eqref{opnaive} is the following. Usually operators with derivatives are power suppressed but in SCET the derivatives corresponding to the large momentum component of the collinear fields are unsuppressed,
\begin{equation}
\bar{n}\cdot \partial\, \phi_c \sim \lambda^0 Q\,  \phi_c\,.
\end{equation}
We therefore need to include operators with an arbitrary number of such derivatives! 

We could work with infinitely many operators and Wilson coefficients $C_n$ for the operator with $n$ such derivatives, but there is a more elegant way to deal with this complication. To understand it, consider the series
\begin{equation}
\phi_c(x+ t \bar{n}) = \sum_{n=0}^\infty \frac{t^n}{n!} (\bar{n} \cdot \partial)^n \phi_c(x)\,.
\end{equation}
Inserting this into a convolution integral we get
\begin{equation}\label{smear}
\int \!dt\, C(t) \phi_c(x+ t n) =  \sum_{n=0}^\infty \frac{C_n}{n!} (\bar{n} \cdot \partial)^n \phi_c(x)\,.
\end{equation}
where $C_n$ is the $n$-th moment of the coefficient function
\begin{equation}
C_n = \int\! dt \,C(t) \, t^n\,.
\end{equation}
Instead of including an arbitrary number of derivatives, we can smear the field $\phi_c$ along the light-cone as in \eqref{smear}. The function $C(t)$ then encodes the information about the Wilson coefficients $a_n$ of the higher-derivative operators.

However, with the smearing the current operator becomes non-local and when putting operators at different points in a gauge theory, we need to be careful to maintain gauge invariance. Consider for example an operator with two collinear fermions at different points
\begin{equation}\label{pdf}
\bar{\xi}_c(x+ t \bar{n}) [ x + t \bar{n}, x ] \frac{ \bar{n}\!\!\!/}{2} \xi_c(x)\,.
\end{equation}
The proton matrix elements of this operator define the quark PDFs. In order to make it gauge invariant one needs to transport the gauge transformation at point $x$ to the point $x+ t \bar{n}$. This can be achieved using the Wilson line
\begin{equation}\label{WilsonStraight}
 [ x + t \bar{n}, x ] = \bm{P} \exp \left[ ig \int_0^t dt' \bar{n}\cdot A_c(x+t \bar{n})\right]\,.
 \end{equation}
Since the exponent is a color matrix, one needs an ordering prescription to define it. The symbol ${\bm P}$ indicates that the matrices at different times should be path ordered, i.e. the one at a later time are to the left of the earlier ones. The Wilson line transforms as
\begin{equation}\label{gaugeWilson}
 [ x + t \bar{n}, x ]  \to  V_c(x+ t \bar{n}) \, [ x + t \bar{n}, x ] \,V^\dagger_c(x) \,,
\end{equation}
which renders the bilocal operator \eqref{pdf} gauge invariant. The gauge transformation \eqref{gaugeWilson} and other properties of Wilson lines are derived in Appendix D of \cite{Becher:2014oda}. Since the other components of the gluon field are power suppressed, a Wilson line along a non-straight path would differ from \eqref{WilsonStraight} by power corrections.

In SCET, it is useful to take a detour and define a Wilson line which runs from infinity along $\bar{n}^\mu$ to the point $x^\mu$ as follows
\begin{equation}
W_c(x) = [ x, x - \infty\, \bar{n} ]\,,
\end{equation}
so that the finite segment can be written as a product 
\begin{equation}
 [ x + t \bar{n}, x ]  = W_c(x+ t \bar{n}) W^\dagger_c(x)\,.
 \end{equation}
 In other words, to move from $x$ to $x+ t \bar{n}$, we first move from $x$  to infinity and then back $x+ t \bar{n}$. The segment we travel in both directions drops out by unitarity of the corresponding matrix, which leaves the finite segment. The advantage of the  Wilson line $W_c$ is that it allows us to define the building blocks
 \begin{equation}\label{build}
 \begin{aligned}
 \chi_c(x) & \equiv W_c^\dagger (x) \xi_c(x) \,,\\
 \mathcal{A}^\mu_c & \equiv W^\dagger_c ( D^\mu_c W_c )\,,
 \end{aligned}
 \end{equation}
 which are invariant under collinear gauge transformations which vanish at infinity. With these building blocks, we can then easily build gauge invariant SCET operators. In addition to the collinear Wilson line $W_c$, we will later also introduce Wilson lines built from soft fields as in \eqref{wilson} to decouple soft interactions.

After all this preparation, we are finally ready to write down the most general leading-power SCET current operator. It takes the form
\begin{equation}\label{currfinal}
J^\mu(0) = \int \!ds \!\int \!dt \,C_V(s,t) \,\bar{\chi}_c(t \bar{n}) \,\gamma_\perp^\mu \, \chi_{\bar{c}}(s n)\,.
\end{equation}
The Wilson coefficient $C_V(s,t)$ (the $V$ stands for vector current) needs to be determined by matching. At tree level it is given by $C_V(s,t)=\delta(t) \delta(s)$ to reproduce the tree-level current \eqref{treeCurrent}. Since it is related to the large derivatives, the Fourier transform of the coefficient encodes the dependence on the large momentum transfer $Q^2 = n\cdot l \, \bar{n}\cdot p$. To see this, we use the momentum operator to shift the fields to the point $x=0$,
\begin{equation}\label{shift}
\phi(x) = e^{i \bm{P} \cdot x} \phi(0) e^{-i \bm{P}\cdot x} \,,
\end{equation}
and take the matrix element of the current operator between a state with an incoming quark $q$ with momentum $l^\mu$ and outgoing one with momentum $p^\mu$. One obtains
\begin{equation}
\begin{aligned}
\langle q(p) | J^\mu(0) | q(l) \rangle
&= \int \!ds \!\int \!dt \,C_V(s,t)\, e^{- i s n\cdot l}\, e^{i t \bar{n}\cdot p}\, \bar{u}(p) \,\gamma_\perp^\mu \, u(l)  \\
&= \tilde{C}_V(n\cdot l \, \bar{n}\cdot p)\,  \bar{u}(p) \,\gamma_\perp^\mu \, u(l)\,.
 \end{aligned}
 \end{equation}
 Note that the Fourier transformed coefficient $\tilde{C}_V$ only depends on the product $Q^2=n\cdot l \, \bar{n}\cdot p$ and not on the individual components. One way to see this is to note that SCET is invariant under a rescaling $n^\mu \to \alpha \,n^\mu$ and $\bar{n}^\mu \to 1/\alpha\, \bar{n}^\mu$. This invariance is manifest already in the decomposition \eqref{momcomp} and is part of a larger set of reparametrization invariances \cite{Manohar:2002fd}, which express the independence of the physics on the exact choice of the reference vectors used to set up the effective theory.

 {\footnotesize
\begin{figure}[t!]        
\begin{center}             
\begin{tabular}{cccc}
\includegraphics[height=0.14\textwidth]{sudakov} & \raisebox{0.4cm}{$=\frac{\alpha_s}{4\pi} \tilde{C}_V^{(1)}(Q^2)$} 
\includegraphics[height=0.14\textwidth]{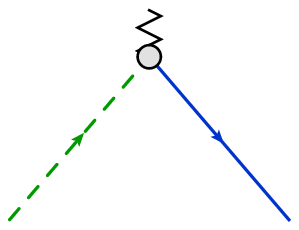} 
 & \raisebox{0.4cm}{$+$} &
 \includegraphics[height=0.14\textwidth]{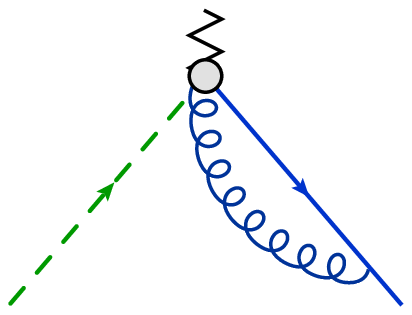}  \\[5pt]
& \raisebox{0.4cm}{$\phantom{=====}+\phantom{=,}$} 
\includegraphics[height=0.14\textwidth]{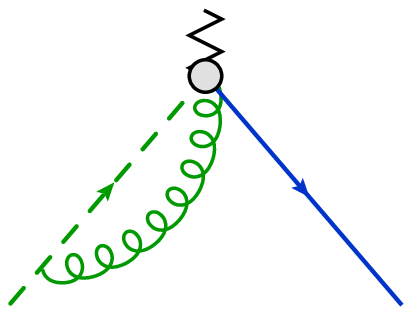} 
 & \raisebox{0.4cm}{$+$} &
\includegraphics[height=0.14\textwidth]{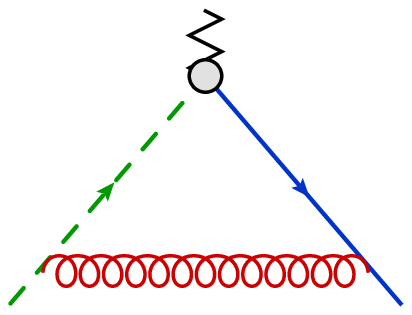} 
\end{tabular}
\end{center}
\caption{One-loop contributions to the Sudakov form factor in QCD and in SCET. In addition to the diagrams shown, there are external leg corrections which we suppress. The Feynman rules for gluons emitted from the current are obtained after expanding the collinear Wilson lines in the fields \eqref{build} the coupling.}  \label{sudakovLoop} 
\end{figure}
}
 
 The Wilson coefficient $\tilde{C}_V$ encodes the dependence on the large momentum scale $Q^2$, which arises from the hard momentum region. To determine it, we should perform a matching computation. The loop diagrams contributing to the one-loop Sudakov form factor are shown in Fig.\ \ref{sudakovLoop}. On the effective field theory side, there are several contributions. In addition to loop diagrams involving collinear fields (blue), anti-collinear fields (green) and soft exchanges (red), there is a contribution from the one-loop correction to the Wilson coefficient. The vertices in which an (anti-)collinear gluon is emitted from the current are obtained after expanding the collinear Wilson lines in the building blocks \eqref{build} in the coupling. The different diagrams are in one-to-one correspondence to the contributions of the different momentum regions computed in Section \ref{RegionsSudakov}, except for the fact that we left out the numerators of the diagrams in the region computation. What remains the same is the fact that each contribution involves a different momentum scale. Setting the low-energy scales $P^2$ and $L^2$ to zero, both the soft and the collinear loop integrals become scaleless. In this case, the full-theory result becomes equal to the contribution of the hard region and on the effective-theory side only the one-loop correction to the Wilson coefficient remains. The most efficient way to extract the Wilson coefficient is thus to compute the on-shell form factor $P^2=L^2=0$. Performing the on-shell computation, one obtains
\begin{equation}
\tilde{C}_V^{\mbox{{\tiny bare}}} (\ep,Q^2)
=  1+ \frac{\alpha_s(\mu)}{4 \pi}
C_F \left(-\frac{2}{\varepsilon^2} - \frac{3}{\varepsilon} +\frac{\pi^2}{6} - 8 + {\mathcal O}(\varepsilon)\right) \left( \frac{Q^2}{\mu^2} \right)^{-\varepsilon}\, + {\mathcal O}\left(\alpha_s^2\right)\,,
\end{equation}
with color structure $t^a t^a = C_F \bm{1}= (N_c^2-1)/(2N_c) \bm{1}$. To get this result, we have expressed the bare coupling $\alpha_s^0=g^2/(4\pi)$ in terms of the $\ms$ renormalized coupling constant $\alpha_s(\mu)$ via the relation $Z_\alpha\, \alpha_s(\mu) \,\mu^{2\ep} = e^{-\ep \gamma_E}(4\pi)^\ep \alpha_s^0$, where $Z_\alpha  = 1+\mathcal{O}(\alpha_s)$ at our  accuracy. We have added a label ``bare'' to the Wilson coefficient to indicate that we still need to renormalize it, which is done by absorbing the divergences into a multiplicative $Z$-factor,
\be\label{eq:renormCV}
\tilde{C}_V(Q^2,\mu) = \lim_{\varepsilon \to 0} Z^{-1}\left(\ep,Q^2, \mu \right) \tilde{C}_V^{\mbox{{\tiny bare}}} (\ep,Q^2) \, .
\ee
Doing so, leaves us with the finite, renormalized Wilson coefficient
\be \label{eq:CVren}
\tilde{C}_V(Q^2,\mu) = 1 + \frac{\alpha_s (\mu)}{4 \pi} C_F \left(
- \ln^2 \frac{Q^2}{\mu^2} +  3 \ln  \frac{Q^2}{\mu^2} + \frac{\pi^2}{6} -8\right)  + {\mathcal O}(\alpha_s^2)\, .
\ee
In the renormalized coefficient we have taken the limit $\ep \to 0$, but it depends on the renormalization scale $\mu$. The whole procedure is of course the same as renormalization in standard quantum field theory, up to the fact that we had to deal with $1/\ep^2$ divergences, which arise because we have both soft and collinear divergences. As a consequence, the Wilson coefficient contains double logarithms. Due to the presence of the double logarithms, the anomalous dimension governing the RG equation for the Wilson coefficient has a logarithmic piece,
\be \label{eq:REGCV}
\frac{d}{d \ln \mu} \tilde{C}_V(Q^2,\mu) = \left[ C_F\, \gamma_{\cusp}(\alpha_s) 
\ln\frac{Q^2}{\mu^2}  + \gamma_V (\alpha_s)\right] \tilde{C}_V (Q^2,\mu) \, ,
\ee
where, at order $\alpha_s$, the functions $\gamma_{\cusp}$ and $\gamma_V$ are given by
\be
\gamma_{\cusp}(\alpha_s)  = 4 \frac{\alpha_s (\mu)}{4 \pi} \, , \qquad
\mbox{and} \qquad \gamma_V (\alpha_s) = - 6 C_F \frac{\alpha_s (\mu)}{4 \pi} \,.
\ee
The on-shell form factor has been computed to three loops \cite{Baikov:2009bg,Gehrmann:2010tu}, and all these ingredients are known to this accuracy. The presence of the logarithm in the anomalous dimension is the distinguishing feature of this RG (and other RG equations in SCET). It is important that only a single logarithm appears so that the expansion of the anomalous dimension is not spoiled by the presence of large logarithms. Otherwise, RG-improved perturbation theory would no longer work. The linearity in the logarithm follows from factorization, which we discuss next.

In soft photon effective theory we were able to decouple the soft radiation from the electron field by the field redefinition \eqref{decoupl} involving a soft Wilson line. In the same way, soft emissions can be decoupled from the collinear field by considering the Wilson line 
\begin{equation}
S_n(x) = \mathbf{P} \exp\left[ ig \int_{-\infty}^0 \!\!ds \, n \cdot A_s(x + s n) \right]\,,
\end{equation}
which fulfills the equation $n\cdot D_s \,S_n(x) = 0$. Redefining
\begin{equation}
\begin{aligned}
\xi_c &= S_n(x_-)\, \xi_c^{(0)} \,,\\
A_c^\mu &= S_n(x_-)\,A_c^{(0)\mu}\,S^\dagger_n(x_-)\,,
\end{aligned}
\end{equation}
the soft-collinear interaction term becomes
\begin{equation}
\begin{aligned}
{\mathcal L}_{c+s} &= \bar{\xi}_c \frac{\nbsl}{2} i n \cdot D \xi_c = \bar{\xi}_c \frac{\nbsl}{2}\left( i n \cdot D_s+ n \cdot A_c \right)\xi_c \\
& =\bar{\xi}^{(0)}_c \frac{\nbsl}{2}\left( i n \cdot \partial_s+ n \cdot A_c^{(0)}\right) \xi^{(0)}_c =  \bar{\xi}^{(0)}_c \frac{\nbsl}{2} i n \cdot D_c^{(0)}\, \xi^{(0)}_c
\end{aligned}
\end{equation}
so that the decoupling has completely removed the soft-collinear interactions from the leading-power Lagrangian. Performing an analogous decoupling also for the anti-collinear fields, it takes the form
\begin{equation}
\mathcal{L}_{{\tiny \mbox{SCET}}} = \mathcal{L}_{c}^{(0)} +  \mathcal{L}_{\bar{c}}^{(0)} + \mathcal{L}_{s}\,.
\end{equation}
Since there are no longer any interactions, we are dealing with independent theories of soft and collinear particles and also the states separate as
\begin{equation}\label{statefact}
| X \rangle = | X_c \rangle\, \otimes\, | X_{\bar{c}} \rangle\, \otimes\, | X_s \rangle\,.
\end{equation}
Of course, this does not imply that the soft physics is no longer present. As in the soft-photon case, it manifests itself as soft Wilson lines along the directions of the energetic particles. The vector current operator $J^\mu$, for example, takes the form
\begin{equation}\label{currFact}
\bar{\chi}_c(t \bar{n}) \,\gamma_\perp^\mu \, \chi_{\bar{c}}(s n) = \bar{\chi}^{(0)}_c(t \bar{n}) \,\bar{S}_n^\dagger(0) \,\gamma_\perp^\mu \, S_{\bar{n}}(0)\, \chi^{(0)}_{\bar{c}}(s n)
\end{equation}
after the decoupling. Note that the decoupling for the anti-collinear fields involves an outgoing soft Wilson line $S_{\bar{n}}$ along the $\bar{n}$ direction, while we needed use the outgoing one for the collinear field $\bar{S}_n$. The soft Wilson lines in the current operator are evaluated at $x=0$, since the position argument in the current has hard scaling because the operator involves both collinear and anti-collinear fields.

{\footnotesize
\begin{figure}[t!bp]        
\begin{center}              
\begin{overpic}[scale=0.7]{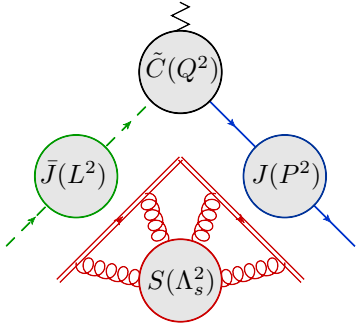}
\put(40,70){$\tilde{C}(Q^2)$}
\put(10,40){$\bar{J}(L^2)$}
\put(70,40){$J(P^2)$}
\put(40.5,10){$S(\Lambda_s^2)$}
\end{overpic}
\end{center}
\caption{Schematic form of the factorized form factor. The red double lines indicate the soft Wilson lines.}   
\label{SudakovFact}    
\end{figure}
}

\section{Resummation by RG evolution}

Computing the factorized form factor in the decoupled theory, we obtain a result of the form
\begin{equation}
F(Q^2,L^2,P^2) = \tilde{C}_V(Q^2,\mu)\, \bar{J}(L^2,\mu)\, J(P^2,\mu) \,S(\Lambda_s^2,\mu)\,,
\end{equation}
which is shown graphically in Fig.\ \ref{SudakovFact}. The collinear and anti-collinear functions $\bar{J}$ and $J$ are of course identical, the bar simply indicates to which sector the function belongs. The soft function is given by the matrix element of the Wilson line in \eqref{currFact} and its scale is $\Lambda_s^2=L^2 P^2/Q^2$. We have indicated that the renormalized effective theory matrix elements and the Wilson coefficient $\tilde{C}_V$ depend on the renormalization scale. This dependence must cancel in the product, which implies that the anomalous dimensions of the ingredients must add up to zero. We gave the RG-equation for the Wilson coefficient in (\ref{eq:REGCV}) and the ones for the collinear and soft factors are
\bea
\frac{d}{d \ln \mu} {J}\left(P^2,\mu^2  \right) &=&
-\left[C_F \,\gamma_{\cusp}\!\left(\alpha_s\right)  \ln{\frac{P^2}{\mu^2}} + \gamma_J\!\left(\alpha_s\right)\right]
{J}\left(P^2,\mu^2  \right) \, , \nn \\
\frac{d}{d \ln \mu} {S}\left(\Lambda_s^2,\mu^2  \right) &=&
\left[C_F \gamma_{\cusp}\!\left(\alpha_s\right) \ln{\frac{\Lambda_s^2}{\mu^2}}  + \gamma_S\!\left(\alpha_s\right)\right]
{S}\left(\Lambda_s^2,\mu^2  \right) \,. 
\eea
The $\mu$-independence of $F(Q^2,L^2,P^2)$ then requires that
\begin{equation}
\begin{aligned}
 &C_F \,\gamma_{\cusp}(\alpha_s) \ln{\frac{Q^2}{\mu^2}} + \gamma_V(\alpha_s) \\
-&C_F\, \gamma_{\cusp}(\alpha_s) \left(\! \ln{\frac{L^2}{\mu^2}} + \ln{\frac{P^2}{\mu^2}} 
\! \right) - 2 \gamma_J(\alpha_s)\\
 +  &C_F \,\gamma_{\cusp}(\alpha_s) \ln{\frac{\Lambda_s^2}{\mu^2}}+ \gamma_S(\alpha_s) = 0 \, .
 \end{aligned}
\end{equation}
 In order for the logarithms to cancel it is crucial that the anomalous dimensions are linear in the logarithm and
 all proportional to the same coefficient $ \gamma_{\cusp}$. To prove linearity one can make a more general ansatz for the anomalous dimensions and then show that scale independence of the sum, together with the kinematic dependence of the individual pieces, imply that higher-log terms must be absent. Note that the soft anomalous dimension is given by a closed Wilson loop with a cusp at $x=0$, where the direction changes from $\bar{n}$ to $n$. Polyakov \cite{Polyakov:1980ca} and Brandt, Neri, and Sato \cite{Brandt:1981kf} proved that Wilson lines with cusps require renormalization and the two-loop anomalous dimension for light-like Wilson lines with a cusp was first computed in \cite{Korchemskaya:1992je}. At the moment, there is a lot of effort to also compute this cusp anomalous dimension to four loops. While the  analytic QCD result is not yet known, some partial analytical and numerical results are available \cite{Lee:2016ixa,Moch:2017uml,Boels:2017skl,Boels:2017ftb,Dixon:2017nat}.
 
 {\footnotesize
\begin{figure}[t!]        
\begin{center}              
\begin{overpic}[scale=0.6]{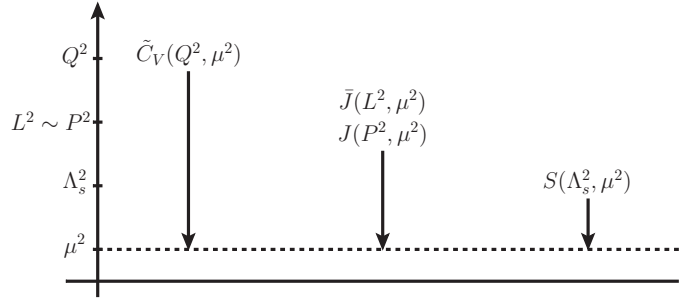}
\end{overpic}
\end{center}
\caption{Resummation by RG evolution. }   
\label{figureRG} 
\end{figure}
}

To resum the large logarithms in the form factor, we can evaluate each ingredient at its characteristic scale and then use the RG to evolve them to a common scale. For the hard function $\tilde{C}_V(Q^2,\mu)$, for example, we chose $\mu_h \sim Q$ as the initial value. For such a choice, the function does not involve any large logarithms and we can evaluate it perturbatively. Solving the RG equation, we then get the hard function also at lower scales as
\begin{equation}
\begin{aligned}
\tilde{C}_V(Q^2,\mu) 
&=  \exp\left\{ \int_{\mu_h}^\mu  d \ln \mu^\prime \left[C_F \gamma_{\cusp}(\alpha_s) 
\ln\frac{Q^2}{\mu^{\prime 2}}  + \gamma_V (\alpha_s) \right] \right\} \, \tilde{C}_V(Q^2,\mu_h)\\
&\equiv U\left(\mu_h,\mu\right) \,\tilde{C}_V(Q^2,\mu_h)\,.
 \end{aligned}
\end{equation}
Since the form of the RG of the other ingredients is the same, also their solution can immediately be written down. Using the definition of the $\beta$-function
\be
\frac{d \alpha_s(\mu)}{d \ln \mu} = \beta\!\left(\alpha_s(\mu) \right) \, 
\ee
and rewriting the logarithm in the exponent of the evolution matrix $U\left(\mu_h,\mu\right)$ as
\be\label{eq:logcoupl}
\ln\frac{\nu}{\mu} =  \int_{\alpha_s(\mu)}^{\alpha_s(\nu)} \frac{d\alpha}{\beta(\alpha)}\, ,
\ee
it can be written in the form 
\be \label{eq:evma}
U\left(\mu_h,\mu\right) = \exp\left[2 C_F S(\mu_h, \mu) - A_{\gamma_V}(\mu_h,\mu) \right] \left( \frac{Q^2}{\mu_h^2}\right)^{- C_F A_{\gamma_{\cusp}} (\mu_h,\mu)} \, .
\ee
The quantities $S$ and $A_\gamma$ are defined as
\bea \label{eq:SAdef}
S\left(\nu,\mu\right) &=& -\int_{\alpha_s(\nu)}^{\alpha_s(\mu)} d \alpha \frac{\gamma_{\cusp} (\alpha)}{\beta(\alpha)} \int_{\alpha_s(\nu)}^{\alpha} \frac{d \alpha'}{\beta(\alpha')} \, ,
\nn \\
A_{\gamma_i}(\nu,\mu)&=& -\int_{\alpha_s(\nu)}^{\alpha_s(\mu)} d \alpha \frac{\gamma_i (\alpha)}{\beta(\alpha)} \, ,
\eea
with $i \in \{V,\cusp\}$. They can be computed by expanding the anomalous dimensions and the $\beta$-function order by order in perturbation theory and then performing the integrations. Their explicit form can be found in the appendix of \cite{Becher:2006mr}. This form of perturbation theory, which involves couplings at the different scales but is free from large logarithms is called RG improved perturbation theory and was covered in Matthias Neubert's lectures \cite{NeubertLecture}. It is the standard way by which resummations are performed in SCET and other effective field theories.

Let me finish this chapter with an admission. While all the steps we took to achieve factorization and to perform resummation would be perfectly appropriate for a physical observable, there is a problem for the  Sudakov form factor. We stated earlier that the off-shell form factor is unphysical and gauge dependent and that we only consider it because it is the simplest example involving both soft and collinear physics. To achieve factorization, we have used the decoupling transformation, which is a field redefinition. Such redefinitions leave physical quantities invariant, as Aneesh Manohar has shown in his lecture \cite{ManoharLecture}, but they do lead to different off-shell Green's functions. In particular, one finds that the soft matrix elements change. After the field redefinition, they involve not only logarithms but also IR divergences and are given by scaleless integrals. The anomalous dimensions we discussed above are not directly affected by this problem, but if one wants to compute them after the decoupling, one will need to separate the UV from the IR divergences. Fortunately, we are not really interested in the off-shell form factor and none of these problems will be present for the physical observables studied in the next chapter.

\chapter{Applications in jet physics}

While the Sudakov form factor discussed in the previous chapter is by itself not a physical quantity, the effective-theory electromagnetic current operator we have constructed is relevant in a variety of physical processes. Three examples are shown in Fig.\ \ref{appls}. On the left, we show the process $e^- + p \to e^- + X$ in a situation, where the hadronic final state $X$ contains many particles. This is called Deep-Inelastic Scattering (DIS) and, as in the case of the Sudakov form factor, SCET is relevant in the limit where the final state invariant mass $M_X$ is much smaller than the momentum transfer $Q$ mediated by the virtual photon. Introducing the variable $x$ via  $M_X^2 = Q^2 (1-x)/x$, this corresponds to the limit $x\to 1$. To analyze the process in SCET, it is easiest to work in the Breit frame in which the virtual photon has momentum $q^\mu = (0,0,0, Q)$. One will then introduce a reference vector $n^\mu$ along the out-going low-mass jet of particles and a vector $\bar{n}^\mu$ for the incoming proton . The vector current of quarks is also relevant at hadron colliders, for example for $p p \to \gamma^*/Z \to e^+ + e^- + X$. Here SCET is relevant in cases where the outgoing radiation $X$ is either soft or collinear to the beam directions and the reference vectors $n^\mu$ and $\bar{n}^\mu$ point along the beams. Finally and most obviously, SCET can be used in cases where there are low mass final-state jets such as the two-jet process displayed on the right side of Fig.\ \ref{appls}. Here the reference vectors point along the jet directions. The processes involving hadrons in the initial state involve collinear matrix elements with protons. Such matrix elements are called {\em beam functions} \cite{Stewart:2009yx}, while the collinear matrix elements with a vacuum initial state are called {\em jet functions}. In addition to a perturbatively calculable part, the beam functions contain the usual PDFs, which arise in all processes with hadrons in the initial state. 

It would of course be interesting to discuss jet production processes at the LHC. However, these processes involve energetic particles both along the beam directions and the jet directions. To analyze two-jet production at the LHC, we would therefore introduce light-cone reference vectors $n_1, \dots , n_4$ and conjugate vectors $\bar{n}_1, \dots , \bar{n}_4$ and one would have four types of collinear fields. To keep things simple, we will stick to processes with only two directions for which we can choose $n_1 = n$ and $n_2 = \bar{n}$. Since we want to have jets in the final state, we will consider leptonic collisions and study the two jet process depicted in the right panel of Fig.\ \ref{appls}. The reader who would like to learn more about PDFs and beam functions can consult the SCET book \cite{Becher:2014oda}, where examples of hadron collider observables  with two energetic directions were discussed, namely threshold resummation and transverse momentum for the Drell-Yan process. 

So far, we have used the word ``jet'' loosely to talk about sprays of energetic particles with low invariant mass. To define jet cross sections, we need to be more concrete. One possibility to define a two-jet process is to demand that all energy except for a small fraction is contained inside two cones. This is the original jet definition proposed by Sterman and Weinberg \cite{Sterman:1977wj} in 1977. This basic idea has evolved into a variety of modern jet definitions, see \cite{Salam:2009jx}. A simpler set of observables are {\em event shapes} which characterize the geometry of collider events. Rather than attributing particles to jets, one introduces a quantity which indicates how pencil-like the final state is. The prototypical event shape is the thrust $T$ introduced by Farhi in the same year as the jet definition \cite{Farhi:1977sg}. 

The resummation of large logarithms for actual jet processes is complicated and most SCET papers have therefore focused on event shapes. We will first analyze the the event shape thrust but will then briefly discuss the resummation for actual jet observables.

{\footnotesize
\begin{figure}[t!]        
\begin{center}              
\begin{tabular}{ccccc}
\raisebox{0.3cm}{\begin{overpic}[scale=0.5]{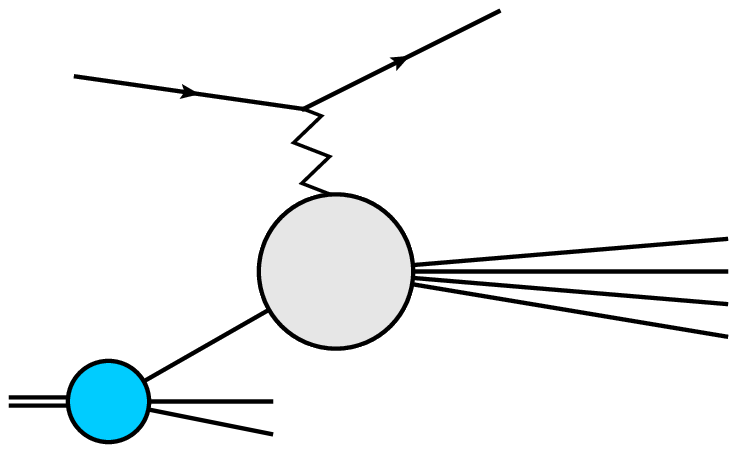}
\put(-10,4){$p$}
\put(-10,50){$e^-$}
\end{overpic}}
&\phantom{sb} & \begin{overpic}[scale=0.5]{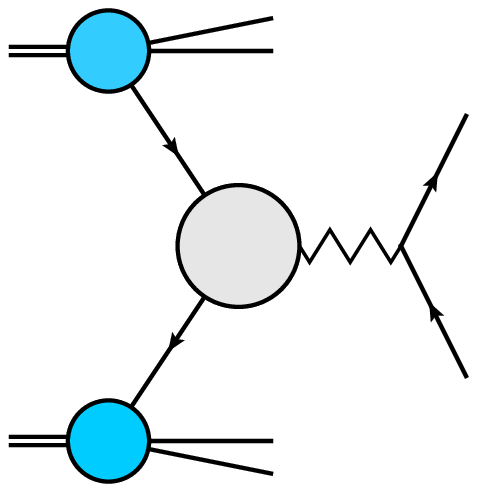}
\put(-10,5){$p$}
\put(-10,88){$p$}
\end{overpic}
& \phantom{sb} & \raisebox{0.2cm}{\begin{overpic}[scale=0.5]{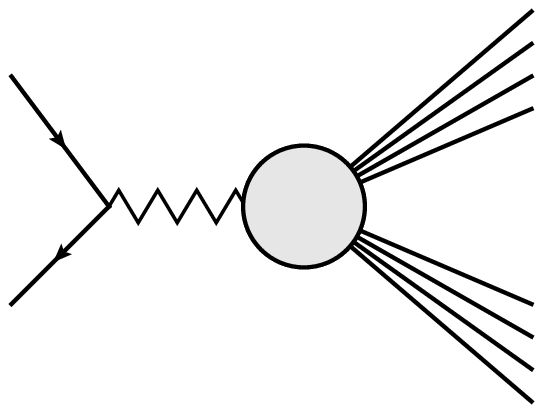}
\put(-14,7){$e^-$}
\put(-14,63){$e^+$}
\end{overpic}}
\end{tabular}
\end{center}
\caption{Collider processes induced the vector or axial-vector current $J^\mu$, indicated by the gray blob. On the left, we have DIS, in the middle the Drell-Yan process and on the right two-jet production in $e^+e^-$ collisions. The processes with protons in the initial state involve PDFs, indicated in blue.}   
\label{appls}    
\end{figure}
}

\section{Factorization for the event-shape variable thrust}

The definition of thrust \cite{Farhi:1977sg},
\begin{equation}\label{thrust}
T  =   \frac{1}{Q} \underset{\vec{n}_T}{{\rm max}} \sum_i | \vec{n}_T \cdot \vec{p}_i |\,,
\end{equation}
needs a few explanations. It involves a sum over all particles in the event and one sums the projections of their momenta along the thrust axis $\vec{n}_T$ which must be chosen to maximize the sum. The momentum flow along $ \vec{n}_T$ is then normalized to the center-of-mass energy, which for massless particles is given by
\begin{equation}
Q = \sum_i | \vec{p}_i |\,.
\end{equation}
The thrust $T$ thus measures the fraction of momentum flowing along the thrust axis. For $T=1$ all momentum must flow along $\vec{n}_T$, i.e. all particles are parallel or anti-parallel to $\vec{n}_T$. Events with large thrust $T$ are thus pencil-like and involve two low-mass jets. It is convenient to define $\tau = 1-T$ such that the end-point is at $\tau=0$. The quantity $\tau$ defines the SCET expansion parameter which was denoted by $\lambda$ in the discussion of the Sudakov problem. The maximum value of $\tau$ (minimum value of $T$) is obtained for a completely spherical event. It is a short exercise to verify that such a configuration has $\tau=1/2$. We show two example events, together with their thrust values in Fig.\ \ref{thrustevent}. 

The definition \eqref{thrust} is useful to distinguish pencil-like from spherical events, but not suitable to single out configurations with more than two jets. However, minimizing over several reference vectors one can generalize thrust in such a way that it vanishes for events with $N$ massless jets. This generalized quantity is called $N$-jettiness \cite{Stewart:2010tn}. At hadron colliders, it is also useful define event shapes in the transverse plane \cite{Banfi:2010xy}. An analysis of factorization for transverse thrust at a hadron colliders can be found in \cite{Becher:2015gsa}. As discussed above, the relevant effective theory involves four collinear sectors.

{\footnotesize
\begin{figure}[t!]        
\begin{center}              
\begin{overpic}[scale=0.5]{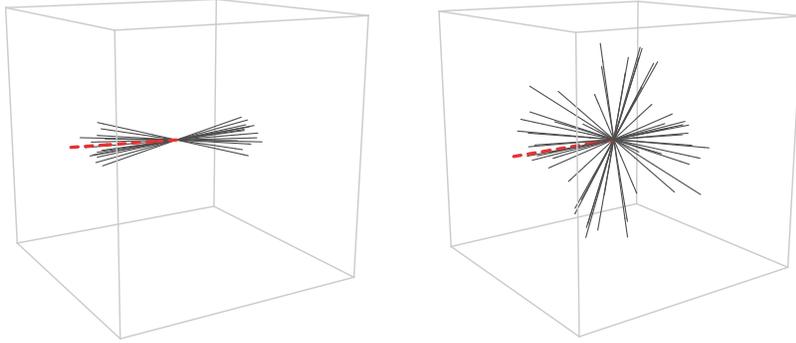}
\end{overpic}
\end{center}
\caption{Two sample collider events. On the left side a pencil-like two-jet event with small $\tau =1-T \approx 0.002$, on the right side an almost spherical event with large $\tau  \approx 0.35$. The thrust axis is shown as a red dashed line.}   
\label{thrustevent}    
\end{figure}
}

Thrust is soft and collinear safe, i.e. its value does not change under exactly collinear splittings or infinitely soft emissions. This property makes it possible to compute it perturbatively. However, for small $\tau \ll 1$ we encounter large logarithms. To analyze this limit, let us choose the SCET reference vectors as $n^\mu = (1,\vec{n}_T)$ and $\bar{n}^\mu = (1,-\vec{n}_T)$. In Fig.\ \ref{largethurst}, we show an event with small $\tau$. It will involve energetic particles collinear to $n^\mu$ and $\bar{n}^\mu$, together with soft large-angle radiation. Performing a region analysis, one will find the same regions 
\eqref{SCETregions} we identified in the Sudakov form factor, but with expansion parameter $\lambda=\tau$. We can thus separate the sum over particles in \eqref{thrust} into individual sums in the soft and collinear sectors and write
\begin{equation}\label{thrustSCET}
\begin{aligned}
\tau Q  &=   \sum_i | \vec{p}_i | - | \vec{n}_T \cdot \vec{p}_i | \\
& =   \sum_i  n\cdot p_{ci}  +  \sum_i  \bar{n}\cdot p_{\bar{c}i}  + \sum_{i}  n\cdot p^R_{si} + \sum_{i}  \bar{n}\cdot p^L_{si} \\
& =  n\cdot p_{X_c} +n\cdot p^R_{X_s} +\bar{n}\cdot p_{X_{\bar{c}}} +\bar{n}\cdot p^L_{X_s} \,,
\end{aligned}
\end{equation}
where we have split the soft particles into left- and right-moving ones in order to be able write the sums in terms of light-cone components. In the last line, we have introduced the total momentum in each category. This result has a simple physical interpretation. Before deriving it, we note that due to the definition of the thrust axis, the total transverse momentum is zero in each hemisphere. As a result, also the {\em total} collinear transverse momentum $p_{X_c}^\perp$ is zero, up to terms which are of the same order as the soft momentum. Up to power corrections, we therefore write the invariant mass of all particles in the right hemisphere as
\begin{equation}
\begin{aligned}
M_R^2 &= (p_{X_c} + p_{X_s}^R)^2 \\
&= p_{X_c}^2 + \bar{n} \cdot p_{X_c}  n \cdot p_{X_s}^R \\
& = \bar{n}\cdot p_{X_c}  n\cdot p_{X_c} + \bar{n} \cdot p_{X_c} n\cdot p_{X_s}^R\\
&= Q (n\cdot p_{X_c} + p_{X_s}^R)\,.
\end{aligned}
\end{equation}
The expression in the last line is exactly the contribution of the right-moving particles to \eqref{thrustSCET}. Up to power corrections, we obtain the equality
\begin{equation}\label{thrustreform}
\tau Q^2 = M_L^2 + M_R^2 = p_{X_c}^2 + p_{X_{\bar{c}}}^2 + Q\left (  n\cdot p_{X_s}^R+  \bar{n}\cdot p_{X_s}^L \right)\, .
\end{equation}
For small values, $\tau$ is equal to the sum of the invariant masses in the two hemispheres, normalized to $Q^2$. The fact that thrust is additive in the soft and collinear contributions, will be important to establish factorization. 

 {\footnotesize
\begin{figure}[t!]        
\begin{center}              
\begin{overpic}[scale=0.6]{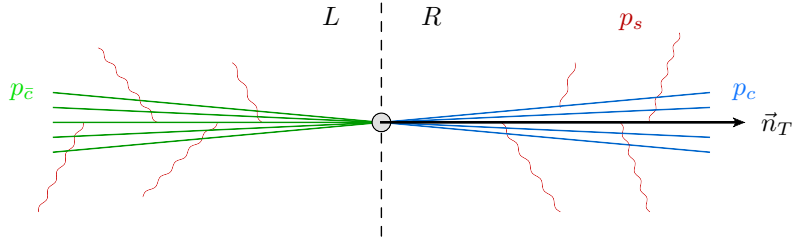}
\put(-4,20){\textcolor{ForestGreen}{$p_{\bar{c}}$}}
\put(98,20){\textcolor{NavyBlue}{$p_{c}$}}
\put(102,15){$\vec{n}_T$}
\put(82,30){\textcolor{BrickRed}{$p_{s}$}}
\put(40,30){$L$}\put(54,30){$R$}
\end{overpic}
\end{center}
\caption{An event with large thrust $T$ (corresponding to small $\tau$). The thrust axis $\vec{n}_T$, shown as an arrow, splits the event into two hemispheres denoted by $L$ and $R$.}   
\label{largethurst}    
\end{figure}
}

Let us now compute the cross section
\begin{equation}
\frac{d\sigma}{d\tau} = \frac{1}{2Q^2} \; \int\limits_{X}\hspace{-0.55cm}\sum \left| \mathcal{M}(e^+e^- \to \gamma^* \to X)\right|^2  (2\pi)^4 \delta^{(4)}(q- p_X)\, \delta( \tau - \tau(X))\,,
\end{equation}
where $\tau$ is the thrust value we prescribe, while $ \tau(X)$ is the value obtained with the given final state momenta in the state $X$, which is equal to the expression in \eqref{thrustreform} at leading power. The momentum $q=q_1+q_2 $ is the total lepton momentum and $q^2=Q^2$. The amplitude can be written as a product of a leptonic amplitude times a hadronic one. For the squared amplitude, this implies $|\mathcal{M}|^2 = L^{\mu\nu} H_{\mu\nu}$, where the hadronic tensor has the form
\begin{equation}\label{Hmunu}
H_{\mu\nu}(q,\tau) =  \int\limits_{X}\hspace{-0.55cm}\sum \langle 0 |\, J^\dagger_\nu(0)\, | X \rangle \langle X |\,  J_\mu(0)\, | 0 \rangle  (2\pi)^4 \delta^{(4)}(q- p_X)\, \delta( \tau - \tau(X))\,,
\end{equation}
and
\begin{equation}\label{Lmunu}
L^{\mu\nu}(q_1,q_2) = \frac{e^4 Q_q^2}{Q^4}\, \bar{v}(q_2)\, \gamma^\mu\, u(q_1) \, \bar{u}(q_1)\, \gamma^\nu \,v(q_2)
\end{equation}
contains the trivial leptonic part. For convenience, we include the photon propagators as well as the quark charge $e Q_q$ in $L^{\mu\nu}$.  Averaging over the spins of the incoming leptons produces a trace of Dirac matrices and the lepton tensor becomes
\begin{equation}\label{Lmunu2}
L^{\mu\nu}(q_1,q_2) = \frac{e^4 Q_q^2}{Q^4}\, \left(q_1^\mu q_2^\nu +q_2^\mu q_1^\nu - q_1\cdot q_2\, g^{\mu\nu} \right)\,.
\end{equation}
We would like the cross section to be differential in the angle $\theta$ of the thrust vector with respect to the incoming electron. To factor out the integral over this direction, we now explicitly distinguish the reference vector $\vec{n}$ in SCET from the thrust axis $\vec{n}_T$, which is derived from the particles in a given event and introduce the dummy integration
\begin{equation}
1 = \int \! d^3 \vec{n}\, \delta^{(3)}(\vec{n} - \vec{n}_T)\,.
\end{equation} 
In the effective theory, the thrust axis is given by $\vec{n}_T = \vec{p}_{X_{c}}/|\vec{p}_{X_{c}}|$ up to power corrections. Inserting this into the above equation and using the fact that  momentum conservation fixes $|\vec{p}_{X_{c}}|=Q/2$, we can rewrite it in the form
\begin{equation}
 \int \! d^3 \vec{n} \,\delta^{(3)}(\vec{n} - \vec{n}_T) = (2\pi) \int d\cos\theta \left(\frac{Q}{2}\right)^2 
 \delta^{(2)}(p_{X_{c}}^\perp)\, ,
\end{equation} 
where the prefactor $(2\pi)$ is from the integration over the azimuthal angle. The net effect of these manipulations is that the cross section involves a $\delta$-function which fixes the total transverse momentum of the collinear radiation to be zero, as required by the thrust definition. Combining it with the momentum conservation $\delta$-functions and expanding away small momentum components, we get
\begin{multline}
 \delta^{(4)}(q- p_{X_c}- p_{X_{\bar{c}}}- p_{X_s})\,\delta^{(2)}\!(p_{X_c}^\perp) \\= 2\, \delta(\bar{n}\cdot p_{X_c}- Q)\,\delta(n\cdot p_{X_{\bar{c}}}- Q)  \delta^{(2)}\!\big(p_{X_c}^\perp\big)  \delta^{(2)}\!\big(p_{X_{\bar{c}}}^\perp\big)\,.
\end{multline}
The factor of 2 is the Jacobian for converting to light-cone components.

After this preparation, we can now plug in the factorized SCET current \eqref{currFact} into the hadron tensor \eqref{Hmunu}. In the Sudakov form factor, we had an incoming and an outgoing particle, so that the current \eqref{currFact} had an incoming soft Wilson line $S_n$ and an an outgoing Wilson line $\bar{S}_{\bar{n}}^\dagger$ after decoupling. In the present case the particle and anti-particle are both outgoing so that the appropriate soft Wilson-lines structure in the current is $\bar{S}_{\bar{n}}^\dagger \bar{S}_n$. Since all QCD particles in this chapter are outgoing, we will drop the bar on the soft Wilson lines in the following.

Using that the states factorize in the form \eqref{statefact}, the hadronic tensor contains  a collinear, an anti-collinear and a soft matrix elements, tied together by the thrust constraint \eqref{thrustreform}. To separate the individual contributions to thrust, we introduce three more integrations
\begin{equation}
\begin{aligned}
1 &= \int \! dM_c^2\, \delta( M_c^2 - p_{X_c}^2) \int \! dM_{\bar{c}}^2\, \delta( M_{\bar{c}}^2 - p_{X_{\bar{c}}}^2)  \int \! d\omega \, \delta( \omega- n\cdot p_{X_s}^R-  \bar{n}\cdot p_{X_s}^L)\,.
\end{aligned}
\end{equation}
Putting all this together, we obtain the cross section in the factorized form
\begin{equation}\label{sigmaFac}
\begin{aligned}
\frac{d\sigma}{d\tau d\cos\theta}= &\frac{\pi}{2}\, L_{\mu\nu}\, |\tilde{C}_V(-Q^2 - i0,\mu)|^2  \int \! dM_c^2\int \! dM_{\bar{c}}^2\,\int \! d\omega \, \delta (\tau -\frac{M_c^2 +M_{\bar{c}}^2+ Q \omega}{Q^2}) \\
&\times\,  \int\limits_{X_c}\hspace{-0.55cm}\sum  \langle 0 | \, {\chi}^a_{c,\delta}(0) | X_c \rangle \langle X_c |\, \bar{\chi}_{c,\alpha}^b\,|0 \rangle\, \delta( M_c^2 - p_{X_c}^2)\, \delta^{(2)}\!\big(p_{X_c}^\perp\big) \delta(\bar{n}\cdot p_{X_c}- Q) \\
&\times\,  \int\limits_{X_{\bar{c}}}\hspace{-0.55cm}\sum  \langle 0 | \, \bar{\chi}^d_{\bar{c},\gamma}(0) | X_{\bar{c}} \rangle \langle X_{\bar{c}} |\, \chi_{\bar{c},\beta}^e\,|0 \rangle\, \delta( M_{\bar{c}}^2 - p_{X_{\bar{c}}}^2)\, \delta^{(2)}\!\big(p_{X_{\bar{c}}}^\perp\big) \delta(n\cdot p_{X_{\bar{c}}}- Q) \\
&\times\, \int\limits_{X_{s}}\hspace{-0.55cm}\sum  \langle 0 | \big[  S_n^\dagger S_{\bar{n}} \big]_{da} \, | X_{s} \rangle \langle X_{s} |\,  \big[  S_{\bar{n}}^\dagger S_n \big]_{be}  \,| 0 \rangle\,  \delta( \omega- n\cdot p_{X_s}^R-  \bar{n}\cdot p_{X_s}^L) \\
& \times  (2\pi)^4 \, (\gamma_\perp^\mu)_{\alpha\beta} \,(\gamma_\perp^\nu)_{\gamma\delta} \,.
\end{aligned}
\end{equation}
In this result, the Latin letters $a, b, d, e$ denote the quark colors, and the Greek letters their Dirac indices. While it is lengthy, the above expression is simply the result of inserting the different ingredients and then separating the contributions to the different sectors. In line two, we have the collinear matrix element defining the jet function $J$,  line three defines $\bar{J}$ and line four the soft function $S$. As they stand, these functions are matrices in color and Dirac space, but the expressions can still be massaged further. An important property is that the collinear matrix elements are color diagonal and therefore proportional to $\delta^{ab} \delta^{de}$. This contracts the color indices in the soft matrix element, and we can define a scalar soft function as
\begin{equation}\label{eq:defSoft}
S(\omega)=\frac{1}{N_c}\, \int\limits_{X_{s}}\hspace{-0.55cm}\sum  \langle 0 | \big[  S_n^\dagger S_{\bar{n}}\big]_{ab} \, | X_{s} \rangle \langle X_{s} |\,  \big[  S_{\bar{n}}^\dagger S_n \big]_{ba} | 0 \rangle\,  \delta( \omega- n\cdot p_{X_s}^R-  \bar{n}\cdot p_{X_s}^L) \,.
\end{equation}
The prefactor $1/N_c$ has been added so that $S(\omega)= \delta( \omega)$ at lowest order. Next, let's consider the collinear and anti-collinear matrix elements in the second and third lines. They can be written in the form
\begin{align}\label{eq:defjetf1}
\frac{\delta^{ab}}{2(2\pi)^3} \left[\frac{{n\!\!\!/}}{2}\right]_{\delta\alpha} \,J(M^2)
&= \int\limits_{X_c}\hspace{-0.55cm}\sum \left<0\right|\chi_{c,\delta}^{a}(0)\left|X_{c}\right>\left<X_{c}\right|\bar{\chi}_{c,\alpha}^{b}(0)\left|0\right> \nonumber \\
&\quad\quad\quad\quad\times \delta( M^2 - p_{X_c}^2)\, \delta^{(2)}\!\big(p_{X_c}^\perp\big) \delta(\bar{n}\cdot p_{X_c}- Q)\,,  \\
\frac{\delta^{de}}{2(2\pi)^3}\left[\frac{\bar{n}\!\!\!/}{2}\right]_{\beta\gamma}  \,J(M^2)
&= \int\limits_{X_{\bar{c}}}\hspace{-0.55cm}\sum\left<0\right|\bar{\chi}_{\bar{c},\gamma}^{d}(0)\left|X_{\bar{c}}\right>\left<X_{\bar{c}}\right|\chi_{\bar{c},\beta}^{e}(0)\left|0\right> \nonumber\\
&\quad\quad\quad\quad\times  \delta( M^2 - p_{X_{\bar{c}}}^2)\, \delta^{(2)}\!\big(p_{X_{\bar{c}}}^\perp\big) \delta(n\cdot p_{X_{\bar{c}}}- Q) \,. \label{eq:defjetf2}
\end{align}
The  final state $X_{c}$ in the collinear matrix element has the quantum numbers of a quark, while the $X_{\bar{c}}$ is a state with anti-quark quantum numbers. This is the case because we have chosen $\vec{n}_T$ to point along the quark direction. To understand why these matrix elements only involve a single scalar jet function $J(M^2)$, we note that due to the property  $n \!\!\!/ \,\chi_{c}=0$, the collinear matrix elements must vanish when multiplied by $n\!\!\!/$ on either side. Because of this constraint the field $\chi_{c}$ is effectively a two-component spinor and the associated Dirac basis has only the four matrices $n \!\!\!/$, $n \!\!\!/\gamma_5$ and $n \!\!\!/ \gamma^\mu_\perp$. Due to parity invariance, the second structure cannot arise and since it does not involve a transverse vector which could be dotted into $\gamma^\mu_\perp$, also the third one must be absent, leaving the first one as the only choice. The prefactor multiplying the Dirac structure on the left-hand side of \eqref{eq:defjetf1} and $\eqref{eq:defjetf2}$ was chosen to have $J(M^2)=\delta(M^2)$ at leading order.

 A convenient representation for computing the jet function is to rewrite the matrix element as the imaginary part of the collinear propagator
\begin{equation}
   J(p^2) = \frac{1}{\pi}\,{\rm Im}\left[ i \mathcal{J}(p^2) \right]
\end{equation}
where
\begin{equation}\label{eq:Jscet}
   \frac{n\!\!\!/}{2}\,\bar{n}\cdot p\, \mathcal{J}(p^2)
   = \int d^4x\,e^{-ip\cdot x}\,\langle 0\,|\,{\rm T}\left\{
   \chi_{c}(0)\,\overline\chi_{c}(x) \right\}|\,0\rangle \,.
\end{equation} 
To derive the equality of the two representations, one inserts a complete set of states into \eqref{eq:Jscet}, then translates the fields from point $x$ to $0$ using \eqref{shift} and uses the Fourier representation 
\begin{equation}
\theta(x^0) = - \frac{1}{2\pi i} \int d\omega \frac{e^{-i \omega x^0}}{\omega + i0}
\end{equation}
for the Heaviside functions in the definition of the time-ordered product. The imaginary parts of the two terms in the time-ordered product then arise from states with quark and anti-quark quantum numbers, as in \eqref{eq:defjetf1} and  \eqref{eq:defjetf2} and are identical by the charge conjugation symmetry of QCD. The representation \eqref{eq:Jscet} makes it clear that the jet function has a simple QCD interpretation. We have stressed earlier that the collinear Lagrangian is equivalent to the QCD Lagrangian and the field $\chi_c = W_c^\dagger P_+ \psi_c$ can be obtained in QCD using the same projection operator and Wilson line. In the light-cone gauge $\bar{n}\cdot A=0$ the Wilson line is absent and we can thus view (and compute \cite{Becher:2010pd}) the jet function as the imaginary part of the quark propagator in this particular gauge.

We can now insert \eqref{eq:defjetf1} and \eqref{eq:defjetf2} into the factorized cross section \eqref{sigmaFac} and use the definition \eqref{eq:defSoft} of the soft function. What remains are some simple manipulations, namely to evaluate the Dirac algebra of the hadron tensor 
\begin{equation}
{\rm tr}\!\left[ \gamma_\mu^\perp \,\frac{\bar{n}\!\!\!/}{2}\,\gamma_\nu^\perp\,\frac{n\!\!\!/}{2}\right] = n_\mu \bar{n}_\nu +n_\nu \bar{n}_\mu -2 g_{\mu\nu}  \equiv  -2 g_{\mu\nu}^\perp
\end{equation}
and to perform the contraction with the leptonic tensor. The contraction produces scalar products with lepton momenta, which can be written in terms of the scattering angle using $n\cdot q_1 = Q/2\,(1+\cos\theta)$, $\bar{n}\cdot q_1 = Q/2\,(1-\cos\theta)$ and the same relations with opposite signs in front of the cosines for $q_2$. Doing so and collecting the prefactors, we arrive at the final result
\begin{align}\label{eq:factttSCET}
\frac{d\sigma}{d\tau d\cos\theta}&=\frac{\pi N_c Q_f^2\alpha^2}{2Q^2}(1+\cos^2\theta) |\tilde{C}_V(-Q^2 - i0,\mu)|^2  \int \! dM_c^2\int \! dM_{\bar{c}}^2\,\int \! d\omega \,  \nonumber\\
&\hspace{2cm}   \delta\Big(\tau -\frac{M_c^2 +M_{\bar{c}}^2+ Q \omega}{Q^2}\Big) J(M_c^2,\mu) \,J(M_{\bar{c}}^2,\mu)\,S(\omega,\mu) \,,
\end{align}
where we give the result for a single quark  flavor with charge $Q_q$ and have written it in terms of the fine-structure constant $\alpha = e^2/(4\pi)$. The dependence on $\theta$ is trivial and we can easily integrate over it. We have defined the jet and soft functions, such that they reduce to $\delta$-functions at lowest order and $\tilde{C}_V= 1 + \mathcal{O}(\alpha_s)$. Plugging in the lowest-order expressions and integrating over $\theta$ and $\tau$ we reproduce the correct lowest order $e^+ e^- \to q \bar{q}$ cross section
\begin{equation}
\sigma_0 = \frac{4 \pi  \alpha ^2 N_c Q_f^2}{3 Q^2}\,.
\end{equation}

\begin{figure}[t!]
 \begin{center}
 \includegraphics[width=\textwidth]{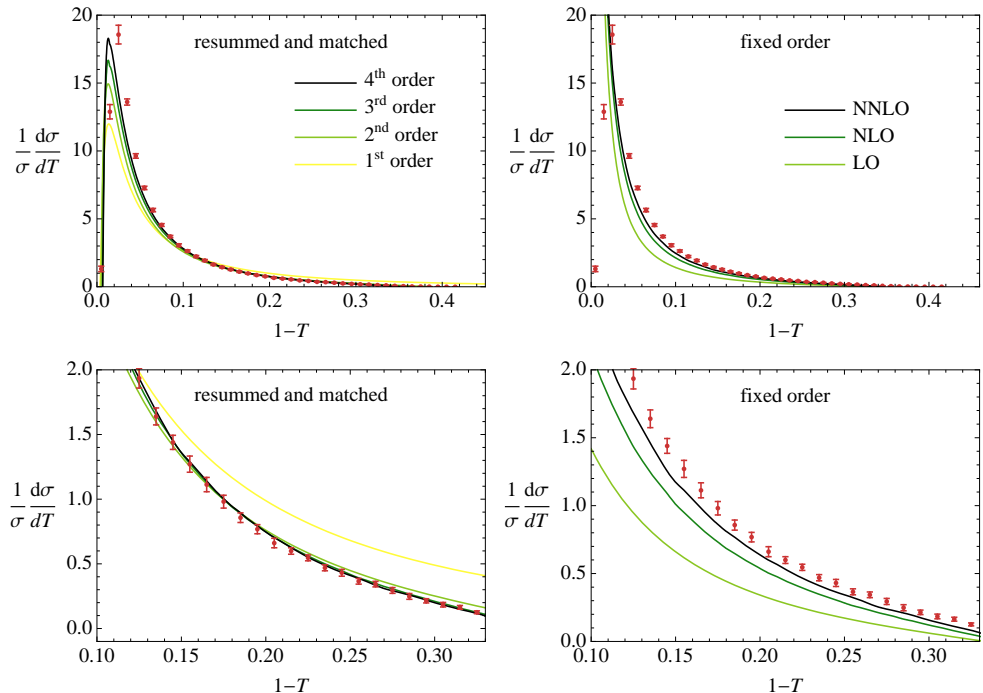}
 \end{center}
 \vspace{-0.3cm}
\caption{Convergence of resummed and fixed-order distributions. {\sc{aleph}} data \cite{Heister:2003aj} (red)
at $91.2$ GeV are included for reference. All plots have $\alpha_s(m_Z)=0.1168$. This figure is taken from \cite{Becher:2008cf}.}
\label{fig:sfconv}
\end{figure}

We have indicated in \eqref{eq:factttSCET} that the hard, jet and soft functions depend on $\mu$. To resum large logarithms, one can again solve the RG equations and evolve to a common reference scale as we discussed for the Sudakov form factor, see Fig.\ \ref{figureRG}.  The only complication is that in the factorization formula \eqref{eq:factttSCET} the soft and jet functions are convolved, while we were dealing with a simple product in the case of the Sudakov form factor. This complication can be avoided by working in Laplace space since the transformation turns the convolution into a product. The Laplace transforms of the jet and soft functions fulfill a RG equation of the same form as $\tilde{C}_V$ and the inversion to momentum space can be performed analytically \cite{Becher:2006nr}. Working in Laplace space, the resummation for thrust was performed at next-to-next-to-leading order (NNLO) in RG improved perturbation theory in \cite{Becher:2008cf}, which corresponds to next-to-next-to-next-to-leading logarithmic (N$^3$LL) accuracy in traditional terminology. For this accuracy one needs the two-loop results for the hard, jet and soft functions, together with three-loop anomalous dimensions. In fact, because of the extra logarithm in the cusp piece, one needs four-loop accuracy for the cusp anomalous dimension $\gamma_{\rm cusp}$; as discussed earlier, the four-loop piece is not yet fully known but has a very small effect on the result. Fig.\ \ref{fig:sfconv}, taken from \cite{Becher:2008cf}, shows the fixed-order expansion to NNLO \cite{GehrmannDeRidder:2007jk}, compared to resummed results at different orders. The plots show the breakdown of fixed-order perturbation theory at small $\tau$, where it diverges, and demonstrate that resummation dramatically improves the convergence of the expansion. We note that the resummed results were matched to the fixed-order predictions at larger $\tau$, i.e. they are constructed in such a way that they reproduce the fixed-order result at higher $\tau$. The different orders in the resummed result shown in the figure correspond to different accuracies in the resummation and in the matching, see \cite{Becher:2008cf} for more details. One-loop calculations for the hard, jet and soft functions can be found in the textbook \cite{Schwartz:2013pla}.

 The resummed result for thrust was fit to the available measurements to extract a value of the strong coupling constant $\alpha_s$ in \cite{Abbate:2010xh}. 
 Since the thrust distribution is affected by nonperturbative hadronization corrections, the authors not only fit for $\alpha_s$ but also for hadronization parameters using data at different center-of-mass energies. The leading effect of hadronization is a shift of the distribution to the right and one indeed observes that the theoretical prediction without hadronization shown in Fig.\  \ref{fig:sfconv} should be shifted to match the experimental result. Fitting for this shift, the authors of \cite{Abbate:2010xh} find significant hadronization effects which reduce the extracted value of $\alpha_s$ by more than $7\%$. To extract $\alpha_s$ reliably, good control over the hadronisation effects and their uncertainty is important.
  The extracted value $\alpha_s(m_Z)=0.1135 \pm (0.0002)_{\rm expt} \pm (0.0005)_{\rm hadr} \pm (0.0009)_{\rm pert}$ from the analysis  \cite{Abbate:2010xh} has small uncertainties and is significantly lower than the world average. The source of this discrepancy is not understood. 
  
\section{Factorization and resummation for jet cross sections}

Interestingly, the pattern of logarithms for jet cross sections such as the Sterman-Weinberg cross section is much more complicated than for thrust. Even at the leading-logarithmic level, one finds new color structures at each loop order rather than a simple exponentiation, as would follow from an RG equation such as \eqref{eq:REGCV} for $\tilde{C}_V$. This complicated structure of logarithms was discovered by Dasgupta and Salam \cite{Dasgupta:2001sh} and arises whenever an observable is insensitive to soft radiation in certain regions of phase space, as is the case for jet cross sections which do not constrain emissions inside the jets. Such observables are called nonglobal and the additional non-exponentiating terms are often called nonglobal logarithms (NGLs). Dasgupta and Salam resummed the leading NGLs at large $N_c$ using a parton shower. They can also be computed by solving a non-linear integral equation derived by Banfi, Marchesini and Smye (BMS) \cite{Banfi:2002hw}. This approach was extended to finite $N_c$ in \cite{Weigert:2003mm} and some finite-$N_c$ results are by now available \cite{Hatta:2013iba,Hagiwara:2015bia}. A resummation of subleading NGLs, on the other hand, has not yet been achieved.

Factorization for such observables was not understood until recently and given that the logarithms obey a non-linear equation, while RG equations are linear, it was even speculated that an effective theory treatment for nonglobal observables might not be possible. One recent proposal to deal with them is to resum global logarithms in a series of subjet observables with the goal of reducing the numerical size of the remaining non-global logarithms after combining the contributions from different subjet multiplicities \cite{Larkoski:2015zka}. At leading-logarithmic accuracy, the expansion in subjects amounts to an iterative solution of the BMS equation \cite{Larkoski:2016zzc}. In a series of recent papers \cite{Becher:2015hka,Becher:2016mmh,Becher:2016omr,Becher:2017nof}, we have attacked the problem directly and have shown that a resumation of NGLs {\em can} be obtained using effective-theory methods. We have derived factorization theorems for a number of such observables, starting with the Sterman-Weinberg cross section. Interestingly, the associated RG equations reduce to a parton shower at leading-logarithmic accuracy. Our effective-theory approach reproduces the leading-logarithmic result of Dasgupta and Salam, but also shows what ingredients are needed to resum subleading logarithms. While they are rather different at first sight, our results closely relate to the color-density matrix formalism of reference \cite{Caron-Huot:2015bja}. We will comment on the precise connection below.

{\footnotesize
\begin{figure}[t!]        
\begin{center}              
\begin{overpic}[scale=0.7]{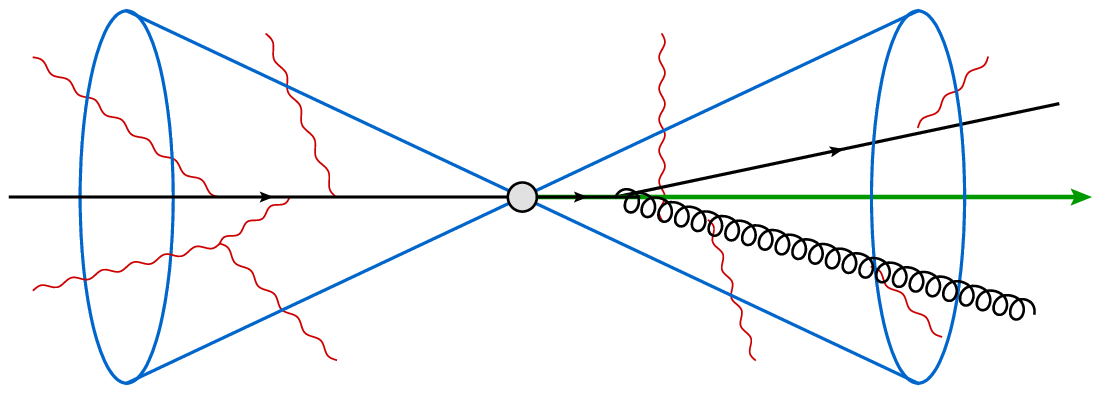}
\put(102,15){\textcolor{ForestGreen}{$\vec{n}_T$}}
\put(35,30){$2 E_{\rm out} < \beta Q$}
\end{overpic}
\end{center}
\caption{A cone jet with three hard partons inside the jets, together with soft radiation.}  
\label{coneJetFig}  
\end{figure}
}

We will now analyze the simplest example of a nonglobal observable, namely the Sterman-Weinberg cross section \cite{Sterman:1977wj}. For simplicity, we choose the thrust axis $\vec{n}_T$ as the jet axis and then put two oppositely directed cones of half-angle $\delta$ on the axis. We will work with a large cone angle $\delta$, and will impose $2E_{\rm out} < \beta Q$ on the energy outside the jet cones. Choosing $\beta\ll 1$ restricts the outside energy to be much smaller than the center-of-mass energy of the collisions $Q$, but the cross section will suffer from large logarithms of $\beta$ which we will resum in the following. The case of small angle $\delta$ was analyzed in \cite{Becher:2015hka,Becher:2016mmh}, but  is more complicated because also collinear logarithms associated with the small opening angle are present and need to be resummed. 
As always, we should first identify which momentum regions are relevant for the problem at hand. This is particularly simple in our setup. Since the opening angle is large, $\delta\sim 1$, there is not really a direction which is singled out. Of course, each event has a thrust axis $\vec{n}_T$, which we use as the jet axis, but since we do not impose that the thrust is large, the particles will in general not be collinear to this axis. So we only have two momentum regions
\begin{equation}\label{JETregions}
\begin{aligned}
&\text{hard} &(h)\quad &  k^\mu \sim Q \,,   \\
&\text{soft} &(s)\quad &   k^\mu \sim \beta Q \,.
\end{aligned}
\end{equation}
The hard momentum region is relevant for the partons inside the jet. These partons cannot be outside otherwise they would violate the constraint $2E_{\rm out} < \beta Q$. Only the soft partons are allowed to be outside the jet. We should now analyze the factorization of the soft partons from the hard ones. Fortunately, and not entirely by coincidence, we have analyzed exactly this situation in the first chapter, where we discussed soft effective theory. We have derived that the soft radiation in QED is given by Wilson lines along the energetic particles, see \eqref{softfact}. The same is true in QCD, except that the soft Wilson lines are color matrices, which act on the colors of the hard amplitude. It is convenient to suppress the color indices and use the color-space notation of Catani and Seymour \cite{Catani:1996jh,Catani:1996vz}, in which the factorization for an amplitude with $m$ hard partons takes the form
\begin{equation}\label{eq:WilsonSoft}
 \bm{S}_1(n_1) \, \bm{S}_2(n_2) \,  \dots\,  {\bm S}_m(n_m)\,|\mathcal{M}_m(\{\underline{p}\})\rangle \,,
\end{equation}
where $\{\underline{p}\}=\{p_1,\dots, p_m\}$ are the particle momenta and $\{\underline{n}\}= \{n_1, \dots , n_m\}$ are light-cone vectors along their directions. The amplitude $|\mathcal{M}_m \rangle$  is a ket vector in color space and the Wilson-line matrix  ${\bm S}_i(n_i)$ acts on the color index of particle $i$. Note that the matrices acting on different partons trivially commute since they act on different indices. To obtain the cross section, we need to square  \eqref{eq:WilsonSoft}, which will give rise to a hard function times a soft matrix element as in \eqref{fact}. However, in Chapter \ref{set} we considered soft radiation for an exclusive process, while we deal with an inclusive cross section in the present case. To get the inclusive result, we need to sum over the number of partons in the final state and integrate over their directions. The cross section then takes the form
\cite{Becher:2015hka,Becher:2016mmh} 
\begin{align}\label{sigbarefinal}
\sigma(Q,\beta) &=  \sum_{m=k}^\infty \big\langle \bm{\mathcal{H}}_m(\{\underline{n}\},Q,\mu) \otimes \bm{\mathcal{S}}_m(\{\underline{n}\},\beta Q,\mu) \big\rangle \,.
\end{align}
The symbol $\otimes$ indicates the integral over the directions and $\langle \dots \rangle$ denotes the color trace, which is taken after multiplying the two functions. The soft function is
\begin{multline}
\bm{\mathcal{S}}_m(\{\underline{n}\},\beta Q,\mu) =\\
 \phantom{aaaaaa}\int\limits_{X_s}\hspace{-0.55cm}\sum \,\langle  0 |\, \bm{S}_1^\dagger(n_1) \,  \dots\,  {\bm S}_m^\dagger(n_m)\,  |X_s \rangle\langle  X_s | \,\bm{S}_1(n_1) \,  \dots\,  {\bm S}_m(n_m) \, |0 \rangle \,  \theta( \beta Q  - 2 E_{\rm \, out}) \,,\label{eq:Sn}
\end{multline}
which can be compared to \eqref{softF} in Chapter \ref{set}. The hard function is
\begin{align}\label{eq:Hm}
\bm{\mathcal{H}}_m(\{\underline{n}\},Q,\mu) =\frac{1}{2Q^2} \sum_{\rm spins}
\prod_{i=1}^m & \int \! \frac{dE_i \,E_i^{d-3} }{(2\pi)^{d-2}} \, |\mathcal{M}_m(\{\underline{p}\}) \rangle \langle  \mathcal{M}_m(\{\underline{p}\}) |\nonumber \\
&\times (2\pi)^d \,\delta\Big(Q - \sum_{i=1}^m E_i\Big) \,\delta^{(d-1)}(\vec{p}_{\rm tot})\,{\Theta }_{\rm in}\!\left(\left\{\underline{p}\right\}\right) \,.
\end{align}
The phase-space constraint ${\Theta }_{\rm in}\!\left(\left\{\underline{p}\right\}\right)$ restricts the hard partons to the inside of the jets. As in our earlier examples, the factorization theorem \eqref{sigbarefinal} achieves scale separation. The hard function only depends on the large scale $Q$, while the soft function depends on the soft scale $Q\beta$. As usual, these functions involve divergences, which can be renormalized after which both the Wilson coefficients $\bm{\mathcal{H}}_m$ and the soft functions $\bm{\mathcal{S}}_m$ depend on the renormalization scale. For experts, let us briefly explain how the above results relate to the formalism of reference \cite{Caron-Huot:2015bja} in which a color-density matrix $U$ is used to track hard partons. The hard functions \eqref{eq:Hm} are obtained by expanding the color-density functional to the $m$-th power in $U$ and  the low-energy Wilson lines matrix elements  \ref{eq:Sn} arise when performing a low-energy average over $U$.

With formula \eqref{sigbarefinal} the resummation of large logarithms (global and nonglobal) becomes standard, at least in principle. All we need to do is to solve the RG evolution for the hard functions
\begin{align}\label{eq:hrdRG}
\frac{d}{d\ln\mu}\,\bm{\mathcal{H}}_m(\{\underline{n} \},Q,\mu)  &= - \sum_{l =k}^{m}  \bm{\mathcal{H}}_l(\{\underline{n} \},Q,\mu) \, \bm{\Gamma}^H_{lm}(\{\underline{n}\},Q ,\mu) \, 
\end{align}
and evolve them from an initial scale $\mu_h \sim Q$, where they are free of large logarithms, down to a lower scale $\mu_s \sim Q\beta$ at which the soft functions are free from large corrections. The only difficulty is that there are infinitely many hard functions $\bm{\mathcal{H}}_l$ which mix under renormalization because the anomalous dimension $\bm{\Gamma}^H_{lm}$ is a matrix not only in color space but also in the multiplicity of the partons. At one loop this matrix has the simple structure
\begin{equation}\label{eq:gammaOne}
\bm{\Gamma} = \frac{\alpha_s}{4\pi}\, \left(
\begin{array}{ccccc}
   \, \bm{V}_{2} &   \bm{R}_{2} &  0 & 0 & \hdots \\
 0 & \bm{V}_{3} & \bm{R}_{3}  & 0 & \hdots \\
0 &0  &  \bm{V}_{3} &  \bm{R}_{3} &   \hdots \\
 0& 0& 0 &  \bm{V}_{4} & \hdots
   \\
 \vdots & \vdots & \vdots & \vdots &
   \ddots \\
\end{array} 
\right)\, + \mathcal{O}(\alpha_s^2)\,,
\end{equation}
because the one-loop correction either are purely virtual in which case they do not change the multiplicity, or involve a single real emission. This structure then imprints itself onto the $\bm{Z}$-factor and the anomalous dimension $\bm{\Gamma}$. That such a mixing must be present is familiar from fixed-order perturbative computations. Amplitudes with fixed multiplicities are not finite and to get a finite result, one needs to combine them with lower multiplicity terms. For example in a NLO computation, the divergences of the real-emission diagrams cancel against the virtual ones. 

The anomalous dimension encodes UV divergences of the soft functions, which are in one-to-one correspondence to soft divergences in the amplitudes. These can be extracted by considering soft limits of amplitudes after which one obtains the explicit results \cite{Becher:2016mmh} 
\begin{align}\label{eq:oneLoopRG}
 \bm{V}_m  &= 2\,\sum_{(ij)}\,(\bm{T}_{i,L}\cdot  \bm{T}_{j,L}+\bm{T}_{i,R}\cdot  \bm{T}_{j,R})  \int \frac{d\Omega(n_l)}{4\pi}\, W_{ij}^l 
 , \\
 \bm{R}_m & = -4\,\sum_{(ij)}\,\bm{T}_{i,L}\cdot\bm{T}_{j,R}  \,W_{ij}^{m+1}\,  \Theta_{\rm in}(n_{m+1})\,, \nonumber
\end{align}
where the dipole radiator is defined as
\begin{equation}
W_{ij}^l = \frac{n_i\cdot n_j}{n_i\cdot n_l \, n_j\cdot n_l}\,.
\end{equation}
This dipole is the combination of two eikonal factors as in \eqref{diagex}, from a soft exchange between legs $i$ and $j$.

{\footnotesize
\begin{figure}[t!]        
\begin{center}              
\begin{overpic}[scale=0.9]{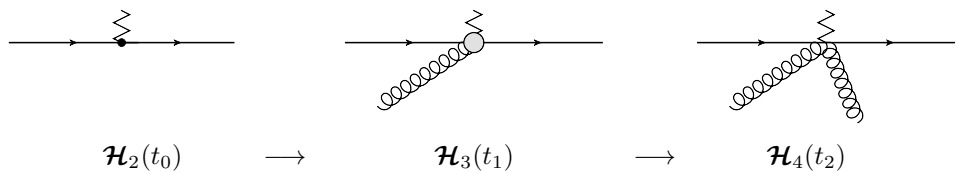}
\put(10,-4){$\bm{\mathcal{H}}_2(t_0)$}
\put(27,-4){$\longrightarrow$}
\put(45,-4){$\bm{\mathcal{H}}_3(t_1)$}
\put(66,-4){$\longrightarrow$}
\put(80,-4){$\bm{\mathcal{H}}_4(t_2)$}
\end{overpic}\\[1cm]
\end{center}
\caption{Solution of the RG equation using Monte Carlo methods.}  
\label{mcFig}  
\end{figure}
}

Let us now consider the resummation of the leading logarithms, which corresponds to evolving with the anomalous dimension at $\mathcal{O}(\alpha_s)$ and evaluating the hard and soft functions at $\mathcal{O}(\alpha_s^0)$. Since the anomalous dimension matrix is very sparse, it is convenient to write out the one-loop RG equation explicitly
\begin{align}\label{eq:diffhrd}
\frac{d}{dt}\,\bm{\mathcal{H}}_m(t)  &=   \bm{\mathcal{H}}_m(t) \,  \bm{V}_m +   \bm{\mathcal{H}}_{m-1}(t) \,  \bm{R}_{m-1} \, ,
\end{align}
where we have traded the dependence on $\mu$ for the variable
\begin{equation}\label{eq:runt}
t = \frac{1}{2\beta_0}\ln\frac{\alpha(\mu)}{\alpha(\mu_h)} =\frac{\alpha_s}{4\pi} \ln\frac{\mu_h}{\mu} +\mathcal{O}(\alpha_s^2)\,,
\end{equation}
which is zero for $\mu=\mu_h$ and increases as we evolve to lower scales. For $\mu = \mu_s$ the logarithm becomes large and compensates the suppression by the coupling constant. For this reason $t$ is treated as a quantity of $\mathcal{O}(1)$ in RG-improved perturbation theory. Solving the homogeneous equation and using variation of the constant, equation \eqref{eq:runt} can also be written as
\begin{equation}\label{eq:MCstep}
\bm{\mathcal{H}}_m(t) = \bm{\mathcal{H}}_m(t_0) \,e^{(t-t_0) \bm{V}_m}
+ \int_{t_0}^{t} dt' \,\bm{\mathcal{H}}_{m-1}(t') \, \bm{R}_{m-1}\, e^{(t-t')  \bm{V}_m}\,.
\end{equation}
This form is usually written for parton showers, which involve an evolution time $t$. One can evolve from
$t_0$ to $t$ either without any additional emissions (first term in \eqref{eq:MCstep}) or by adding an emission to the lower multiplicity cross section (second term). The connection to parton showers becomes even more clear when we consider the initial condition. At the high scale $\mu= \mu_h$, corresponding to $t=0$, only the hard function $\bm{\mathcal{H}}_2$ is present since the higher multiplicity functions involve powers of $\alpha_s$ and are free of large logarithms for this scale choice. Starting with the function $\bm{\mathcal{H}}_2$, we can then iteratively generate the higher functions as
\begin{align}\label{MC}
\bm{\mathcal{H}}_2(t) &= \bm{\mathcal{H}}_2(0) \,e^{t \bm{V}_2}\,, \nonumber\\
\bm{\mathcal{H}}_{3}(t) &= \int_{0}^{t} dt' \,\bm{\mathcal{H}}_{2}(t') \, \bm{R}_{2}\, e^{(t-t')  \bm{V}_{3}} \,,\\
\bm{\mathcal{H}}_{4}(t) &= \int_{0}^{t} dt' \,\bm{\mathcal{H}}_{3}(t') \, \bm{R}_{3}\, e^{(t-t')  \bm{V}_{4}} \,,\nonumber \\
\bm{\mathcal{H}}_{5}(t) &= \dots\,, \nonumber 
\end{align}
see Figure \ref{mcFig}. To get the resummed result, one evolves to the appropriate value of $t$, which is set by the scales $\mu_h$ and $\mu_s$ in \eqref{eq:runt}. The leading-logarithmic cross section is obtained from the sum
\begin{align}\label{eq:sigmaLL}
\sigma_{\rm LL}(Q,\beta) &= \sum_{m=2}^\infty \big\langle  \bm{\mathcal{H}}_m(t) \,\hat{\otimes}\, \bm{1} \big\rangle \nn \\
&= \big\langle \bm{\mathcal{H}}_2(t) + \int \frac{d\Omega_1}{4\pi} \bm{\mathcal{H}}_{3}(t) +\int \frac{d\Omega_1}{4\pi}\int \frac{d\Omega_2}{4\pi}  \bm{\mathcal{H}}_{4}(t) + \dots \big\rangle \,,
\end{align}
where we have used that the soft functions at the scale $\mu_s$ are trivial at leading order $\bm{\mathcal{S}}_m = \bm{1}+\mathcal{O}(\alpha_s)$. The integrals over $t$ as well as the angular integrals for the cross section can be computed using Monte-Carlo methods as is done in all parton shower programs. To do so, one generates a time step $\Delta t_1$, evaluates $\bm{\mathcal{H}}_2(t_1)$ at $t_1=\Delta t_1$ using \eqref{MC}. This is then used as the initial condition for computing $\bm{\mathcal{H}}_{3}(t_2)$ after the next time step $\Delta t_2$ which yields $t_2= \Delta t_1 +\Delta t_2$ after randomly choosing the direction of the extra parton, etc. Repeating these steps one obtains an ensemble of hard functions which yield the cross section \eqref{eq:sigmaLL}. The only stumbling block for an implementation is that the hard functions and anomalous dimensions are matrices in color space. Also note that the shower acts on the level of the amplitudes in \eqref{eq:Hm}, not on the cross section. In fact, our shower is within the general class of showers described in \cite{Nagy:2007ty}. To avoid the difficulty with color, one can take the large $N_c$ limit which renders the color structure trivial, see \cite{Becher:2016mmh} for details. An efficient implementation of the resulting parton shower was given by Dasgupta and Salam in \cite{Dasgupta:2001sh}. 

It is remarkable that we recover a parton shower as the solution of a RG equation in SCET. Of course it is not a general purpose shower, but simply computes the logarithms for a class of observables and nothing more,  but also nothing less. We are guaranteed that we obtain the correct logarithmic structure from this shower, since it is based on a factorization theorem in QCD. The NGL shower differs in some important aspects from standard Monte-Carlo showers. For example, these codes take great care to conserve momentum in the emissions, while our shower does not do this, since small momenta are systematically expanded away. We only generate angles, not full momenta, because the evolution generates the logarithms of the energy. However, since our shower is based on RG equations, we know which elements are needed to extend the resummation to higher logarithms, in contrast to standard showers. Next-to-leading logarithmic resummation corresponds to NLO in RG improved perturbation theory. One will thus need to include the hard functions and soft functions to $\mathcal{O}(\alpha_s)$. Concretely, this implies that the one-loop corrections to $\bm{\mathcal{H}}_{2}(0)$ and the tree-level $\bm{\mathcal{H}}_{3}(0)$, together with the one-loop correction to all functions $\bm{\mathcal{S}}_m$ need to be included, as well as the two-loop anomalous dimension. This anomalous dimension, based on soft limits, was computed in the color-density matrix formalism in \cite{Caron-Huot:2015bja} (for $\mathcal{N}=4$ even three-loop corrections were obtained \cite{Caron-Huot:2016tzz}). It should therefore be possible to set up a shower that also performs subleading resummation for nonglobal observables.

The connection to parton showers is also interesting from the point of view of automating resummations. For global observables automated resummation was pioneered in \cite{Banfi:2004yd} at NLL and extended to NNLL for leptonic collisions in \cite{Banfi:2014sua}. Also in the framework of SCET some resummations have been automated. These include NNLL resummation for processes involving a jet-veto \cite{Becher:2014aya}, NLL resummation for two-jet event shapes \cite{Farhi:2015jca} and the NNLO computation of soft functions \cite{Bell:2015lsf,Bell:2018jvf}. Such automation will be important to make higher-logarithmic computations more available and to extend them to more complicated observables, similarly to what has been achieved by automating fixed-order computations up to NLO.

Given that our field of research is high-energy physics, an effective theory for energetic particles has obviously many applications, beyond the few examples discussed in these lectures. An overview over the areas in which this effective field theory has been used can be found in the last chapter of the book \cite{Becher:2014oda}. In addition to collider-QCD applications, there has been a lot of work on heavy-quark physics (in particular on $B$-meson decays), work on electroweak resummation for collider processes and dark-matter annihilation, work on low-energy QED processes, applications in heavy-ion collisions, as well as more exotic topics such as soft-collinear gravity \cite{Beneke:2012xa,Okui:2017all} and supersymmetry \cite{Cohen:2016dcl}. An important area of recent progress concerns Glauber gluons. A version of SCET which includes their effects has been put forward \cite{Rothstein:2016bsq} and been used to study factorization violation \cite{Schwartz:2017nmr,Schwartz:2018obd}. Another important recent application of SCET has been to use it not just for resummations, but to simplify fixed-order computations. Using a method called $N$-jettiness subtraction \cite{Boughezal:2015dva,Gaunt:2015pea}, a generalization of $q_T$-subtraction \cite{Catani:2007vq}, one is able to trade an NNLO computation in QCD for an NNLO computation in SCET, together with an NLO computation in QCD. This technique has successfully been used to perform many NNLO computations. Jet substructure (see \cite{Larkoski:2017jix} for a recent review) is another area where there has been a lot of progress using SCET. Factorization theorems for jet substructure observables have been derived \cite{Larkoski:2015kga,Larkoski:2017cqq} and substructure techniques have been used to improve observables such that they can be predicted with better accuracy \cite{Larkoski:2015zka,Frye:2016aiz,Hoang:2017kmk}. Finally, we mentioned earlier that the analysis of power corrections has progressed a lot over the past year \cite{Moult:2016fqy,Boughezal:2016zws,Bonocore:2016awd,Feige:2017zci,Moult:2017jsg,Goerke:2017lei,Beneke:2017ztn}. These examples illustrate that while SCET has become a a standard method to analyze factorization and perform resummation, the development of these methods continues to be a very active area of research.

\acknowledgements

I would like to thank the school organizers Sacha Davidson, Paolo Gambino, Mikko Laine and Matthias Neubert for the opportunity to present these lectures. I thank the students for their attention, for interesting discussions, and of course the wine. I attended a Les Houches summer school as a PhD student and it was a great pleasure to be back (and slightly shocking to discover how many years passed since my first visit). Two of the students, Marcel Balsiger and Philipp Schicho were kind enough to read through a draft version of these lecture notes (they have the bad fortune to do their PhD here in Bern so that I could easily solicit their help). I thank them for pointing out some of the more egregious typos; please let me know if you find more! I thank Dingyu Shao and Rudi Rahn for discussions, and Andrea Ferroglia and Alessandro Broggio for letting me recycle some equations from our book.

\end{document}